\documentclass[showpacs,preprint,preprintnumbers,showpacs,showkeys,superscriptaddress,amsmath,amssymb,nofootinbib]{revtex4-2}
\usepackage{color}
\usepackage{latexsym}
\usepackage{amsmath}
\usepackage{amssymb}
\usepackage{eufrak}
\usepackage{euscript}
\usepackage[latin1]{inputenc}
\usepackage{pstricks}
\usepackage{epstopdf}
\usepackage{epsfig}
\usepackage{graphics}
\usepackage{graphicx}
\usepackage{picture}
\usepackage{bbold}
\usepackage{bm}
\usepackage{appendix}

\newcommand{\be}{\begin{equation}}
\newcommand{\ee}{\end{equation}}
\newcommand{\ba}{\begin{eqnarray}}
\newcommand{\ea}{\end{eqnarray}}

\begin{document}

\title{\Large{The axion-photon mixing in non-linear electrodynamic scenarios}}

\author{ J. M. A. Paix\~ao} \email{jeff@cbpf.br}
\affiliation{Centro Brasileiro de Pesquisas F\'isicas, Rua Dr. Xavier Sigaud
150, Urca, CEP 22290-180, Rio de Janeiro, Brazil}
\author{ L. P. R. Ospedal } \email{leonardo.ospedal@ufrgs.br }
\affiliation{Centro Brasileiro de Pesquisas F\'isicas, Rua Dr. Xavier Sigaud
150, Urca, CEP 22290-180, Rio de Janeiro, Brazil}
\affiliation{ Instituto de F\'isica, Universidade Federal do Rio Grande do Sul,
Av. Bento Gon\c{c}alves 9500, CEP 91501-970, Porto Alegre, Brazil}
\author{M. J. Neves}\email{mariojr@ufrrj.br}
\affiliation{Departamento de F\'isica, Universidade Federal Rural do Rio de Janeiro, BR 465-07, CEP 23890-971, Serop\'edica, RJ, Brazil}
\author{J. A.  Helay\"el-Neto}\email{helayel@cbpf.br}
\affiliation{Centro Brasileiro de Pesquisas F\'isicas, Rua Dr. Xavier Sigaud
150, Urca, CEP 22290-180, Rio de Janeiro, Brazil}

\begin{abstract}
In this contribution, we re-assess some aspects of axionic electrodynamics by coupling non-linear electromagnetic effects to axion physics. We present a number of motivations to justify the coupling of the axion to the photon in terms of a general non-linear extension of the electromagnetic sector. Our emphasis in the paper relies on the investigation of the constitutive permittivity and permeability tensors, for which the axion contributes by introducing dependence on the frequency and wave vector of the propagating radiation. Also, we point out how the axion mass and the axion-photon-photon coupling constant contribute to a dispersive behavior of the electromagnetic waves, in contrast to what happens in the case of non-linear extensions, when  effective refractive indices appear which depend only on the direction of the propagation with respect to the external fields. The axion changes this picture by yielding refractive indices with dependence on the wavelength. We apply our results to the special case of the (non-birefringent) Born-Infeld Electrodynamics and we show that it becomes birefringent whenever the axion is coupled. The paper is supplemented by an Appendix, where we follow our own path to approach the recent discussion on a controversy in the definition of the Poynting vector of axionic electrodynamics.

\end{abstract}

\maketitle

\newpage

\pagestyle{myheadings}
\markright{The axion-photon mixing in non-linear electrodynamic scenarios}
%
%
%
\section{Introduction}
\label{sec:1}
Axions are actively investigated in the literature ever since their proposal by Peccei and Quinn to solve the problem of strong CP-violation \cite{Peccei,Peccei2}. More generally, inspired by axionic QCD, Axion-like Particles (ALPs) are treated as pseudo Nambu-Goldstone bosons that arise in various extensions of the Standard Model (SM), and are promising candidates for a dark matter portion in the universe \cite{Wilczekaxion,Sikivie,Dine}. Unlike of the axionic QCD, in which axions couple with the gluons, the scalar ALPs $(\phi)$ couple with the photon through the interaction
$g_{a\gamma} \, \phi \, \left({\bf E}\cdot{\bf B}\right)$, where $g_{a\gamma}$ is a coupling constant with length dimension. Several efforts have been made in an attempt to detect these particles, whether in astrophysical observations or in terrestrial experiments such as particle accelerators, or high-intensity lasers. It is important to emphasize the huge range of possibilities for the mass of the ALPs. In this sense, there are two perspectives in the search for ALPs, the first one takes into account the scattering processes that are capable of producing ALPs in the mass range $\mbox{eV}-\mbox{TeV}$.
The second one considers astrophysical observations, where it is capable to produce ALPs with an upper bound for the mass
which may be in the range of $10^{-10} \, \mbox{eV} - 30 \, \mbox{KeV}$ \cite{Ayala}.
For instance, the Chandra's data analysis for the active galactic nucleus NGC 1275 at the center of the Perseus cluster provides the most stringent limit on the ALP-photon coupling constant for very light ALPs, {\it i.e.}, $g_{a\gamma} < \left(6.3-7.9\right) \times 10^{-13} \,   \mbox{GeV}^{-1}$ for $m_a < 10^{-12} \, \mbox{eV}$ depending on the magnetic field, at $99.7\% $ confidence level \cite{Reynolds}. Recent searches for ALP-Photon resonant conversion on magnetar SGR J1745-2900 exclude couplings $g_{a\gamma} > 10^{-12} \, \mbox{GeV}^{-1}$ for $m_a \leq 10^{-6} \, \mbox{eV}$ \cite{Bondarenko}. Another well-established limit was obtained in CAST, which searches for axions coming from the solar core by converting the X-rays into axions via a magnetic field up to $9.5 \, \mbox{T}$. They report the upper limit on the $g_{a\gamma} \simeq 0.66 \times 10^{-10} \, \mbox{GeV}^{-1}$ for $m_a < 0.02 \, \mbox{eV}$ at 95\% confidence level \cite{CAST}.
For larger values of the ALPs masses, we must take into account the bounds obtained by experiments in particle accelerators. As example, axions can be produced in the reaction $\gamma\gamma \to \phi \to \gamma\gamma\;$ through the Primakoff-process at the LHC. In this case, there is a wide range for the  ALPs masses, ranging from $\mbox{eV}$ to $\mbox{TeV}$ scale \cite{SCHOEFFEL,Baldenegro1,Baldenegro2}. For lead-lead collisions, the exclusion limits for the ALP masses is $m_a \simeq (5-100)\, \mbox{GeV}$, for a coupling constant of $g_{a\gamma} \simeq 0.05 \, \mbox{TeV}^{-1}$ at 95 \% confidence level \cite{d'Enterria}. Since ALPs are searched for in experiments with intense magnetic fields, it is reasonable to expect that non-linear effects might be excited in this situation. They may arise whenever magnetic fields are close to the critical Schwinger magnetic field, $|{\bf B}|_S = m_e^2/q_e = 4.41 \times 10^{9}\,\mbox{T}$ \cite{Schwinger}. In astrophysical searches for ALPs, there are cases of blazars and magnetars with magnetic fields of the order or higher than the critical Schwinger magnetic field \cite{Duncan,jacobsen2022constraining,Bondarenko}. In a recent work, it has been discovered that non-linearities can also arise at low magnetic fields in a QED level for the so-called Dirac materials. In this case, strong non-linear effects arise in dirac materials for a magnetic field strength of approximately $1\,\mbox{T}$ \cite{2022nonlinear}. Several non-linear theories are candidates to extension of the Maxwell electrodynamics (ED) in the literature \cite{Plebanski,Boillat,Birula1,Birula2,Sorokin,TTdeformations1,TTdeformations2}. One of the most known is the Born-Infeld (BI) ED that was originally proposed to remove the singularity of the electric field of a point-like charged particle at the origin \cite{Born}. It is worth mentioning that BI electrodynamics was investigated in a scenario involving axions, in which the analog of Snell's law was found considering an axionic domain wall \cite{colegarussa}. Currently, the BI ED emerges in scenarios of superstring theory, quantum-gravitational models and magnetic monopoles \cite{Fradkin,Pope,Banerjee,Mann,Garcia,NiauEPJC,NiauPRD}. The measurement of the light-by-light scattering at the ATLAS Collaboration of the LHC imposes a lower bound $\gtrsim 100$ GeV on the BI parameter \cite{Ellis}. More stringent bounds on the BI parameter are also discussed in the electroweak model with the hypercharge sector associated with the BI model \cite{MJNevesPat}.
It is known that the vacuum can be considered as a non-linear optical medium and that this concept applies to the standard model of elementary particles \cite{Battesti}. In this sense, non-linear ED models may present dichroism and vacuum birefringence phenomena. In particular, the vacuum magnetic birefringence (VMB) is a macroscopic quantum effect predicted by QED \cite{EulerHeisen,Weisskopf} in which the difference of the refractive indices between parallel and perpendicular polarized light is non-trivial in the presence of an external magnetic field. The PVLAS (Polarisation of Vacuum with LASER) experiment carried out 25 years of efforts in the search for the vacuum birefringence and dichroism, and although it did not reach the values predicted by the QED, it established the best limits known so far \cite{Zavattini,25years,DellaValle,Zavattini_Universe}. Although the VBM has not yet been directly detected, the indirect evidence has been found in the neutron star $\mbox{RX}\; \mbox{J}1856.5 - 3754$, with magnetic fields on the order of $10^{13}\; \mbox{Gauss (G)}$ \cite{Mignani}. It is interesting to notice that the VBM phenomenon can be a tool for the detection of ALPs, where it is known that the conversion of photons into axions changes the polarization of the incident beam by means of a magnetic background field. Thereby, a measure of the birefringence can provide important limits on the axion mass and the coupling constant $g_{a\gamma}$ \cite{Maiani,Yamazaki}.
In this direction, we point out that the birefringence effects related to the axion field in the presence of a laser beam were investigated in ref. \cite{Villalba}. We also highlight that the birefringence phenomena associated with a pure electric background field is due to the optical Kerr effect \cite{Scott,Rizzo}.

At this stage, it should be mentioned that other extensions of the Maxwell electrodynamics (ED) coupled to the axion field have been investigated in the literature. For instance, in a seminal paper by Raffelt and Stodolsky \cite{Raffelt_1988}, the authors included the non-linear effects of the Euler-Heisenberg electrodynamics and obtain the correspondent dispersion relations of the non-linear photon. We also highlight that the axion field theory was considered in connection with high-order derivative Podolsky ED \cite{Gaete_1}, where the effective photonic theory and the inter-particle potential were carried out. Similarly, the axion field contributions were investigated in the context of non-commutative field theories \cite{Gaete_2,Gaete_3}. Furthermore, in the work of ref. \cite{Li_Cheng_2008}, by using the Proca theory, the authors analyzed the influence of a massive photon and its effects on the axion-photon mixing. Later, the axion-Proca ED was obtained as an effective field theory in a Condensed Matter system \cite{Chrispim}. Recently, new extensions involving a hidden photon (another massive dark matter candidate) coupled to the axion and photon fields have been proposal in the literature \cite{Huang_Lee,Arias}. The propagation effects in the presence of extra CPT-odd terms also motivate other extensions of the Maxwell ED \cite{PedroSilvaPRD2021}. The study of the constitutive relations on the wave propagation in bi-isotropic and anisotropic media has applications in material physics \cite{PedroSilvaArxiv2022}. Axionlike couplings can be generated via quantum corrections in a Lorentz violation background \cite{BorgesPRD2014}.

In this paper, we propose the study of a general non-linear ED coupled to an axionic scalar field, that we call $\phi$, coupled
to the non-linear sector through the interaction $g \, \phi \, \left({\bf E}_{0}\cdot{\bf B}_{0}\right)$. We start the description of the
model with a general non-linear kinetic sector. We expand the $4$-potential associated with electromagnetic (EM) fields $({\bf E}_{0},{\bf B}_{0})$
around an  EM background up to second order in the propagating EM fields. Thereby, we have a general linearized ED propagating in an EM background field coupled to an axionic scalar field. We explore the propagation effects for the plane wave solutions in which the dispersion relations, the group velocities, the electric permittivity, and the magnetic permeability are obtained in a uniform and constant EM background fields. The Born-Infeld (BI) non-linear theory is considered as an application of these results. Therefore, we discuss the properties of the wave propagation, like the dispersion relations,
group velocities and the characteristics of the medium in the presence of an external magnetic field, and posteriorly,
of an external electric field. We analyze the results in a regime of strong magnetic field for the BI theory. The birefringence phenomena is also investigated in the BI theory for the cases with magnetic and electric background fields, separately. In the case of the birefringence with a magnetic background, we calculate the axion coupling constant using the data of the PVLAS-FE experiment.
For the birefringence with an electric background, we make a connection with the optical Kerr effect.

The paper is organized with the following outline. In Section \ref{sec2},
we describe the non-linear ED-axion model in an EM background, and obtain the corresponding field equations for the axion and photon.
Next, Section \ref{sec3} focus on the properties of plane wave solutions and we organize our results in two subsections :
the first subsection \ref{subsec3A} considers the purely magnetic background case. The second subsection \ref{subsec3B} discusses a purely electric background. In Section \ref{sec4}, we apply all the results of the previous Section to BI ED, for a magnetic background in subsec. \ref{subsec4A}; next, we go into a purely electric background in subsec. \ref{subsec4B}. After that, in Section \ref{sec5}, we investigate the
birefringence of the axion-BI model for a wave in a magnetic (subsec. \ref{subsec5A}) and in an electric background (subsec. \ref{subsec5B}).
Section \ref{sec6} casts our Concluding Comments. Finally, we include an Appendix \ref{appendix}, where
the energy-momentum tensor of a general non-linear ED and the corresponding conserved quantities are obtained in a constant and uniform EM background.
Throughout this paper, we adopt natural units $\hbar=c=1$  with $4 \pi \epsilon_0 = 1$, and the Minkowski
metric $\eta^{\mu\nu}=\mbox{diag}(+1,-1,-1,-1)$. The electric and magnetic fields have squared-energy mass dimension in which the conversion of Volt/m and Tesla (T) to the natural system is as follows: $1 \, \mbox{Volt/m}=2.27 \times 10^{-24} \, \mbox{GeV}^2$ and $1 \, \mbox{T} =  6.8 \times 10^{-16} \, \mbox{GeV}^2$, respectively.

%
\section{Non-linear electrodynamics coupled to an axionic scalar field}
\label{sec2}
We start up with the Lagrangian (density) of the model
\begin{eqnarray}\label{Lmodel}
{\cal L}={\cal L}_{nl}({\cal F}_{0},{\cal G}_{0})
+\frac{1}{2} \, \left(\partial_{\mu}\phi \right)^{2}
-\frac{1}{2} \, m^2 \, \phi^2
+ g \, \phi \, {\cal G}_{0}-J_{\mu}\,A^{\mu} \, \; ,
\end{eqnarray}
where $\mathcal{L}_{nl}({\cal F}_{0},{\cal G}_{0})$  denotes the most general Lagrangian of a non-linear electrodynamics that is function of the Lorentz- and gauge-invariant bilinears
$
{\cal F}_{0}=-\frac{1}{4} \, F_{0\mu\nu}^{2} ,
$
and
$
{\cal G}_{0}=-\frac{1}{4} \, F_{0\mu\nu}\widetilde{F}_{0}^{\;\,\,\mu\nu} .
$
We consider the antisymmetric field strength tensor as $F_{0}^{\;\,\,\mu\nu}=\partial^{\mu}A_{0}^{\;\,\nu}-\partial^{\nu}A_{0}^{\;\,\mu}=\left( \, -E_{0}^{\;\,i} \, , \, -\epsilon^{ijk}B_{0}^{\;\,k} \, \right)$,
and the correspondent dual tensor is $\widetilde{F}_{0}^{\;\,\,\mu\nu}=\epsilon^{\mu\nu\alpha\beta}F_{0\alpha\beta}/2=\left( \, -B_{0}^{\;\,i} \, , \, \epsilon^{ijk}E_{0}^{\;\,k} \, \right)$, which satisfies the Bianchi identity $\partial_{\mu}\widetilde{F}_{0}^{\;\,\mu\nu}=0$. Therefore, in terms of electromagnetic fields, the invariants are ${\cal F}_{0} = \frac{1}{2} \, \left( {\bf E}_{0}^2-{\bf B}_{0}^2\right)$ and ${\cal G}_{0} = {\bf E}_{0}\cdot{\bf B}_{0}$. In addition, $\phi$ corresponds to the axion  field with mass $m$, and $g$ is the non-minimal
coupling constant (with length dimension) of the axion with the electromagnetic field, {\it i.e.}, the usual coupling with the ${\cal G}_{0}$-invariant. There are many investigations and experiments to constraint the possible regions in the space of the parameters $g$ and $m$, which still remains with a wide range of values, depending on the phenomenological scale in analysis. For more details, we point out the recent reviews \cite{Pierre,PDG}.

The action principle related to the Lagrangian (\ref{Lmodel}) leads to the motion equations with {\color{red} a} classical external current $J^{\nu}$ :
\begin{subequations}
\begin{eqnarray}\label{eqs}
\partial_{\mu}\left(\frac{\partial {\cal L}_{nl}}{\partial{\cal F}_{0}} \, F_{0}^{\;\,\mu\nu}+\frac{\partial {\cal L}_{nl}}{\partial{\cal G}_{0}} \, \widetilde{F}_{0}^{\;\,\mu\nu}\right)&=&-g \,  \left( \partial_{\mu}\phi \right) \, \widetilde{F}_{0}^{\;\,\mu\nu}+J^{\nu} \; ,
\\
\left(\Box+m^2\right)\phi&=&g \, {\cal G}_{0}
\; ,
\end{eqnarray}
\end{subequations}
in which the current is conserved $\partial_{\mu}J^{\mu}=0$.

We expand the abelian gauge field as $A_{0}^{\;\,\mu}=a^{\mu}+A_{B}^{\;\;\,\,\mu}$,  with $a^{\mu}$ being the photon $4$-potential,
and $A_{B}^{\;\;\,\,\mu}$  denotes a background potential. In this conjecture, the tensor $F_{0}^{\;\,\,\mu\nu}$ is also written as the
combination $F_{0}^{\;\,\,\mu\nu}=f^{\mu\nu} \, + \, F_{B}^{\;\;\,\mu\nu}$, in which $f^{\mu\nu}=\partial^{\mu}a^{\nu}-\partial^{\nu}a^{\mu}=\left( \, -e^{i} \, , \, -\epsilon^{ijk}b^{k} \, \right)$ is the EM field strength tensor that propagates in the space-time, and $F_{B}^{\;\;\,\mu\nu}=\left( \, -E^{i} \, , \, -\epsilon^{ijk}B^{k} \, \right)$ corresponds to the EM background field. The notation of the $4$-vector and tensors with index
(B) indicates that it is associated with the background. At this stage, we consider the general case in
which the background depends on the space-time coordinates. Under this prescription, we also expand the Lagrangian (\ref{Lmodel})
around the background up to second order in the propagating field $a^{\mu}$ to yield the expression
\begin{eqnarray} \label{L4}
 {\cal L}^{(2)}  &=&  -\frac{1}{4} \, c_{1} \, f_{\mu\nu}^{\, 2}
-\frac{1}{4} \, c_{2} \, f_{\mu\nu}\widetilde{f}^{\mu\nu}
-\frac{1}{2} \, G_{B\mu\nu} \, f^{\mu\nu}
+\frac{1}{8} \, Q_{B\mu\nu\kappa\lambda} \, f^{\mu\nu}f^{\kappa\lambda}
+\frac{1}{2} \, \left(\partial_{\mu}\phi \right)^{2}
-\frac{1}{2} \, m^2 \, \phi^2
\nonumber \\
&-&  \frac{1}{2} \, g \, \phi \, \widetilde{F}_{B\mu\nu} \, f^{\mu\nu}
+ g \, \phi \, \mathcal{G}_{B}
-J_{\mu}\,a^{\mu}-J_{\mu}\,A_{B}^{\;\,\,\,\mu}+{\cal L}_{nl}\left({\cal F}_{B},{\cal G}_{B}\right) \; ,
\end{eqnarray}
where the background tensors are defined by
\begin{eqnarray}
G_{B\mu\nu}&=& c_{1} \, F_{B\mu\nu}+  c_{2} \, \widetilde{F}_{B\mu\nu} \; ,
\nonumber \\
Q_{B\mu\nu\kappa\lambda} &=& d_{1} \, F_{B \mu\nu} \, F_{B\kappa\lambda}
+d_{2} \, \widetilde{F}_{B \mu\nu} \, \widetilde{F}_{B \kappa\lambda}
+d_{3} \, F_{B \mu\nu}\widetilde{F}_{B \kappa\lambda}
+ d_{3} \, \widetilde{F}_{B \mu\nu} F_{B \kappa\lambda} \; ,
\end{eqnarray}
and ${\cal L}_{nl}\left({\cal F}_{B},{\cal G}_{B}\right)$ is the non-linear Lagrangian as function of the Lorentz invariants
${\cal F}_{B}=-\frac{1}{4} \, F_{B\mu\nu}^2={\bf E}^2-{\bf B}^2$ and ${\cal G}_{B}=-\frac{1}{4} \, F_{B\mu\nu}\widetilde{F}_{B}^{\;\;\,\mu\nu}={\bf E}\cdot{\bf B}$, both in terms of the EM background field, and $\widetilde{F}_{B}^{\;\,\,\mu\nu}=\epsilon^{\mu\nu\alpha\beta}F_{B\alpha\beta}/2=\left( \, -B^{i} \, , \, \epsilon^{ijk}E^{k} \, \right)$ is the dual tensor of $F_{B\mu\nu}$. Furthermore, the coefficients $c_{1}$, $c_{2}$, $d_{1}$, $d_{2}$ and $d_{3}$ are evaluated at ${\bf E}$ and ${\bf B}$, as follows :
\begin{eqnarray}\label{coefficients}
c_{1}&=&\left.\frac{\partial{\cal L}_{nl}}{\partial{\cal F}_{0}}\right|_{{\bf E},{\bf B}}
, \,
\left. c_{2}=\frac{\partial{\cal L}_{nl}}{\partial{\cal G}_{0}}\right|_{{\bf E},{\bf B}}
, \,
\left. d_{1}=\frac{\partial^2{\cal L}_{nl}}{\partial{\cal F}_{0}^2}\right|_{{\bf E},{\bf B}}
, \,
\left. d_{2}=\frac{\partial^2{\cal L}_{nl}}{\partial{\cal G}_{0}^2}\right|_{{\bf E},{\bf B}}
, \,
\left. d_{3}=\frac{\partial^2{\cal L}_{nl}}{\partial{\cal F}_{0}\partial{\cal G}_{0}}\right|_{{\bf E},{\bf B}} ,
\hspace{0.4cm}
\end{eqnarray}
that depend on the EM field magnitude and it may also be functions of the space-time coordinates.
Following the previous definitions of the tensors, we have that $G_{B\mu\nu}=-G_{B\nu\mu}$, and $Q_{B\mu\nu\kappa\lambda}$ is symmetric under exchange $\mu\nu \leftrightarrow \kappa\lambda$, and antisymmetric when $\mu \leftrightarrow \nu$ or $\kappa \leftrightarrow \lambda$.  Note that the current $J^{\mu}$ also couples to the external potential $A_{B}^{\;\,\,\,\mu}$, but this term and ${\cal L}_{nl}\left({\cal F}_{B},{\cal G}_{B}\right)$ are irrelevant for the field equations in which we are interested. If we consider the scalar potential as $V(\phi)= m^2\phi^2/2 - g \, \phi \, {\cal G}_{B}$, it has a minimal at $\phi_{0}=g \, {\cal G}_{B}/m^2$. Writing $\phi = \widetilde{\phi} + \phi_{0}$, the term $g\,\phi\,{\cal G}_{B}$ can be eliminated in the Lagrangian (\ref{L4}) :
\begin{eqnarray}\label{L4phitil}
 {\cal L}^{(2)}  &=&  -\frac{1}{4} \, c_{1} \, f_{\mu\nu}^{\, 2}
-\frac{1}{4} \, c_{2} \, f_{\mu\nu}\widetilde{f}^{\mu\nu}
-\frac{1}{2} \, H_{B\mu\nu} \, f^{\mu\nu}
+\frac{1}{8} \, Q_{B\mu\nu\kappa\lambda} \, f^{\mu\nu} \, f^{\kappa\lambda}
+\frac{1}{2} \, (\partial_{\mu}\widetilde{\phi})^{2}
-\frac{1}{2} \, m^2 \, \widetilde{\phi}^2
\nonumber \\
&-&  \frac{1}{2} \, g \, \widetilde{\phi} \, \widetilde{F}_{B\mu\nu} \, f^{\mu\nu}
-J_{\mu}\,a^{\mu}-J_{\mu}\,A_{B}^{\;\,\,\,\mu}+{\cal L}_{nl}\left({\cal F}_{B},{\cal G}_{B}\right)+ \frac{g^2 \, {\cal G}_{B}^2}{2m^2}
\;,
\end{eqnarray}
where $H_{B\mu\nu} = G_{B\mu\nu}+ g^2\,\mathcal{G}_B \, \widetilde{F}_{B\mu\nu}/m^2$. In this context, the scalar field $\widetilde{\phi}$ is reinterpreted as the axion field with mass $m$. It should be noted that $\phi_{0}$ is non-trivial only in the presence of both electric and magnetic background fields.
Using the action principle in relation to $a^{\mu}$, the Lagrangian \eqref{L4phitil} yields the EM field
equations with source $J^{\mu}$
\begin{eqnarray}\label{EqGmunu}
 \partial^\mu \left[ c_1 \, f_{\mu \nu} + c_2 \, \widetilde{f}_{\mu \nu} -
\frac{1}{2} \, Q_{B\mu \nu \kappa \lambda} \, f^{\kappa \lambda} \right] = - g \, ( \partial^\mu \widetilde{\phi} ) \,
\widetilde{F}_{B\mu \nu}  - \partial^\mu H_{B \mu \nu} + J_\nu \, \; \, ,
\end{eqnarray}
and the Bianchi identity remains the same one for the photon field, namely, $\partial_{\mu}\widetilde{f}^{\mu\nu}=0$.
The action principle related to $\widetilde{\phi}$ in eq. (\ref{L4phitil}), now yields the axion field equation evaluated
at the EM background :
 \begin{eqnarray}\label{eqescalar}
\left(\Box+m^2\right)\widetilde{\phi}=
-\frac{1}{2} \, g \, \widetilde{F}_{B\mu\nu} \, f^{\mu\nu}
\; .
\end{eqnarray}
Notice that, when we fix $c_1=1$ and $d_{1}=d_{2}=d_{3}=0$, the non-linear effects disappear, and we have the usual Maxwell ED coupled to the axion field and  EM background. In the limit $g \rightarrow 0$, the axion uncouples the photon field, and we have a simple combination of a massive free scalar field with the Maxwell ED. Moreover, the Maxwell equations also are recovered in eq. (\ref{EqGmunu}) by taking the aforementioned considerations and turn-off the background fields, $F_{B\mu\nu}=0$.
\section{The dispersion relations in the presence of magnetic and electric  background fields}
\label{sec3}
In this section, we obtain the dispersion relations (DRs) associated with the axion and photon fields in the presence of a uniform and constant electromagnetic background. Thereby, from now on, all the coefficients defined in eq. (\ref{coefficients}) are not dependent on  the space-time coordinates. We start with the equations written in terms of the fields ${\bf e}$ and ${\bf b}$. For the study of the wave propagation, we just consider the linear terms in ${\bf e}$, ${\bf b}$ and $\widetilde{\phi}$, as well as the equations with no current and source, ${\bf J}= \bf{0}$ and $\rho=0$. Under these conditions, the modified electrodynamics from eq. (\ref{EqGmunu}) and Bianchi identity is read below :
\begin{subequations}
\begin{eqnarray}
\nabla\cdot{\bf D}&=&0 \;\;\; , \;\;\;
\nabla\times{\bf e}+\frac{\partial {\bf b}}{\partial t} = {\bf 0} \; ,
\label{divDrote}
\\
\nabla\cdot{\bf b}&=&0 \;\;\; , \;\;\;
\nabla\times{\bf H}-\frac{\partial {\bf D}}{\partial t} = {\bf 0} \; ,
\label{divbrotH}
\end{eqnarray}
\end{subequations}
where the vectors ${\bf D}$ and ${\bf H}$ are given by
\begin{subequations}
\begin{eqnarray}
{\bf D}&=&c_{1} \, {\bf e}
+ d_{1} \, {\bf E} \, ({\bf E}\cdot{\bf e})
+d_{2} \, {\bf B} \, ({\bf B}\cdot{\bf e})-d_{1} \, {\bf E} \, ({\bf B}\cdot{\bf b})
+d_{2} \, {\bf B} \, ({\bf E}\cdot{\bf b})+g \, \widetilde{\phi} \, {\bf B} \, , \;\;\;\;
\label{D}
\\
{\bf H}&=&c_{1} \, {\bf b}
- d_{1} \, {\bf B} \, ({\bf B}\cdot{\bf b})
-d_{2} \, {\bf E} \, ({\bf E}\cdot{\bf b})+d_{1} \, {\bf B} \, ({\bf E}\cdot{\bf e})
-d_{2} \, {\bf E} \, ({\bf B}\cdot{\bf e})-g \, \widetilde{\phi} \, {\bf E} \, . \;\;\;\;
\label{H}
\end{eqnarray}
\end{subequations}
Observe that, in eqs. (\ref{D}) and (\ref{H}), we have eliminated the terms with dependence on the coefficient $d_{3}$, since $d_{3}=0$
in the non-linear ED model in which we will consider ahead. The axion field equation (\ref{eqescalar})  leads to
\begin{eqnarray}\label{EqscalarEB}
\left(\Box+m^2\right)\widetilde{\phi}=
g \left({\bf e}\cdot{\bf B}\right)
+g \left({\bf b}\cdot{\bf E}\right) \; .
\end{eqnarray}
We write the Fourier integrals for the fields ${\bf e}$, ${\bf b}$ and $\widetilde{\phi}$  such that
the field equations (\ref{divDrote}), (\ref{divbrotH}) and (\ref{EqscalarEB}) in momentum space are given by :
\begin{subequations}
\begin{eqnarray}
{\bf k} \cdot {\bf D}_0 &=& 0
\;\;\; , \;\;\;
{\bf k} \times {\bf e}_0 - \omega \, {\bf b}_0 = {\bf 0} \; ,
\label{kD0ktimese0}
\\
{\bf k} \cdot {\bf b}_0 &=& 0
\;\;\; , \;\;\;
{\bf k} \times {\bf H}_0 +\omega \, {\bf D}_0  = {\bf 0} \; ,
\label{kb0ktimesH0}
\\
\left({\bf k}^2-\omega^2+m^2\right)\widetilde{\phi}_0 &=& g \left({\bf B} \cdot {\bf e}_{0}\right)
+g \left({\bf E} \cdot {\bf b}_{0}\right)
\; ,
 \label{axionphi0}
\end{eqnarray}
\end{subequations}
where the amplitudes $D_{0i}$ and $H_{0i}$ are functions of the ${\bf k}$-wave vector and the frequency $\omega$. In terms of the electric and magnetic amplitudes $e_{0i}$ and $b_{0i}$, we obtain
\begin{subequations}
\begin{eqnarray}
D_{0i}({\bf k},\omega) &=& \varepsilon_{ij}({\bf k},\omega)\,e_{0j}+\sigma_{ij}({\bf k},\omega)\,b_{0j}
\; ,
\label{D0i}
\\
H_{0i}({\bf k},\omega) &=& -\sigma_{ji}({\bf k},\omega)\,e_{0j}+ (\mu^{-1})_{ij}({\bf k},\omega)\,b_{0j}
\; ,
\label{H0i}
\end{eqnarray}
\end{subequations}
in which the electric permittivity symmetric tensor $\varepsilon_{ij}({\bf k},\omega)$ and
$\sigma_{ij}({\bf k},\omega)$ are defined by
\begin{subequations}
\begin{eqnarray} \label{eij}
\varepsilon_{ij}({\bf k},\omega) &=& c_{1} \, \delta_{ij} + d_1 \, E_{i} \, E_{j}
+ d_2 \, B_{i} \, B_{j}+ \frac{g^2 \, B_{i} \, B_{j}}{{\bf k}^2-\omega^2+m^2} \; ,
\\
\sigma_{ij}({\bf k},\omega) &=& -d_{1} \, E_{i} \, B_{j} + d_{2} \, B_{i} \, E_{j}
+ \frac{g^2 \, B_{i} \, E_{j}}{{\bf k}^2-\omega^2+m^2} \; .
\end{eqnarray}
\end{subequations}
In addition, $\mu^{-1}$ stands for the inverse of the magnetic permeability symmetric tensor, with the components
\begin{eqnarray}\label{muijinv}
(\mu^{-1})_{ij} = c_{1} \, \delta_{ij} - d_1 \, B_{i} \, B_{j} - d_2 \, E_{i} \, E_{j} - \frac{g^2 \, E_{i} \, E_{j}}{{\bf k}^2-\omega^2+m^2} \; .
\end{eqnarray}
The inverse of eq. (\ref{muijinv}) yields the following expression for the magnetic permeability
\begin{equation} \label{muij}
\mu_{ij}({\bf k},\omega)= \frac{1}{c_1} \frac{\left(1-d_B\,{\bf B}^2-d_{E}\,{\bf E}^2\right)\delta_{ij}+d_B\, B_{i}\,B_{j}+d_{E}\,E_{i}\,E_{j}+d_{B}\,d_{E}\left({\bf E}\times{\bf B}\right)_{i}\left({\bf E}\times{\bf B}\right)_{j}}{1-d_B\,{\bf B}^2-d_{E}\,{\bf E}^2+d_{B} \, d_{E} \, ({\bf E}\times{\bf B})^{2}} \; ,
\end{equation}
where we adopted the shorthand notations
\begin{eqnarray}
d_{B}:=\frac{d_1}{c_1}
\hspace{0.5cm} \mbox{and} \hspace{0.5cm}
d_{E}:=\frac{d_2}{c_1}+\frac{g^2/c_1}{{\bf k}^2+m^2-\omega^2} \; ,
\end{eqnarray}
for simplicity of the equations. In both the cases in which ${\bf E}={\bf 0}$ or ${\bf B} = {\bf 0}$, we have $\sigma_{ij}({\bf k},\omega)=0$. Moreover, it is important to note that the dependence of electric permittivity and magnetic permeability on ${\bf k}$ and $\omega$ is exclusively due to the presence of the axion coupling, see eqs. (\ref{eij}) and (\ref{muij}), as well as the above definition for the coefficient $d_{E}$.

The eigenvalues of the electric permittivity matrix are given by
\begin{subequations}
\begin{eqnarray}
\lambda_{1\varepsilon} &=& c_1 \; ,  \\
\lambda_{2\varepsilon} \! &=& \! c_1\left(1 + \frac{d_B\,{\bf E}^2}{2} + \frac{d_{E} \, {\bf B}^2}{2}\right)
-c_1 \, \sqrt{ \, \left(\frac{d_B\,{\bf E}^2}{2} - \frac{d_{E}\,{\bf B}^2}{2} \right)^2
\!\!+d_{B}\,d_{E}\left({\bf E}\cdot{\bf B}\right)^2} \; ,   \\
\lambda_{3\varepsilon} &=& c_1\left(1 + \frac{d_B\,{\bf E}^2}{2} + \frac{d_{E} \, {\bf B}^2}{2}\right)
+c_1 \, \sqrt{ \, \left(\frac{d_B\,{\bf E}^2}{2} - \frac{d_{E}\,{\bf B}^2}{2} \right)^2
\!\!+d_{B}\,d_{E}\left({\bf E}\cdot{\bf B}\right)^2 } \; .
\end{eqnarray}
\end{subequations}
The correspondent eigenvectors are known as the optical axes of the system.
If these eigenvalues are positive, it satisfy the conditions
\begin{equation}\label{condepsilon}
c_{1} > 0
\hspace{0.3cm} \mbox{and} \hspace{0.3cm}
1+d_{E}\,{\bf B}^2+d_{B} \, {\bf E}^2 + d_{B} \, d_{E} \left( {\bf E} \times {\bf B} \right)^{2} > 0 \; ,
\end{equation}
and the electric permittivity matrix will be defined positive.
The eigenvalues of the magnetic permeability matrix are
\begin{subequations}
\begin{eqnarray}
\lambda_{1\mu} &=& \frac{1}{c_1}  \; ,
\\
\lambda_{2\mu} &=& \frac{1}{2c_1} \frac{2-d_{B}\,{\bf B}^2-d_{E}\,{\bf E}^2-\sqrt{\left(d_{B} \, {\bf B}^2-d_{E} \, {\bf E}^2\right)^{2}+ 4\,d_{B} \, d_{E} \left({\bf E}\cdot{\bf B}\right)^{2} } }{1-d_B\,{\bf B}^2-d_{E}\,{\bf E}^2+d_{B} \, d_{E} \, ({\bf E}\times{\bf B})^{2}} \; ,
\\
\lambda_{3\mu} &=& \frac{1}{2c_1} \frac{2-d_{B}\,{\bf B}^2-d_{E}\,{\bf E}^2+\sqrt{\left(d_{B} \, {\bf B}^2-d_{E} \, {\bf E}^2\right)^{2}+ 4\,d_{B} \, d_{E} \left({\bf E}\cdot{\bf B}\right)^{2} } }{1-d_B\,{\bf B}^2-d_{E}\,{\bf E}^2+d_{B} \, d_{E} \, ({\bf E}\times{\bf B})^{2}}  \; ,
\end{eqnarray}
\end{subequations}
where the permeability is positive if we impose the conditions
\begin{eqnarray}\label{condmu}
c_{1} > 0 \hspace{0.3cm} \mbox{and} \hspace{0.3cm} 1-d_{B}\,{\bf B}^2-d_{E} \, {\bf E}^2 + d_{B} \, d_{E} \left( {\bf E} \times {\bf B} \right)^{2}>0 \; .
\end{eqnarray}
However, situations with negative eigenvalues are also acceptable and this would indicate, according to the references \cite{livroCaloz,Veselago,Padilla}, that the vacuum manifests the behaviour of a category of metamaterial.
If we just consider the magnetic background field $({\bf E}=0)$, the electric permittivity has two degenerated eigenvalues,  $\lambda_{1\varepsilon}=\lambda_{2\varepsilon}$. Analogously, if the background is purely electric $({\bf B}={\bf 0})$,
the magnetic permeability has the two degenerated eigenvalues, $\lambda_{1\mu}=\lambda_{2\mu}$.
In the limit  $g \rightarrow 0$, the conditions (\ref{condepsilon}) and (\ref{condmu})
are reduced to $c_1^2+c_1\,d_{1}\,{\bf B}^2+c_1\,d_{2} \, {\bf E}^2 + d_{1} \, d_{2} \left( {\bf E} \times {\bf B} \right)^{2} > 0$
and $c_1^2-c_1\,d_{1}\,{\bf B}^2-c_1\,d_{2} \, {\bf E}^2 + d_{1} \, d_{2} \left( {\bf E} \times {\bf B} \right)^{2} > 0$, respectively.
For the particular case of Maxwell ED coupled to the axion field in which $c_1=1$ and $d_{1}=d_{2}=0$ in eqs. (\ref{condepsilon}) and (\ref{condmu}), we arrive at the following constraints on the frequency : $-\sqrt{{\bf k}^2+m^2+g^2\,{\bf E}^2}<\omega<\sqrt{{\bf k}^2+m^2+g^2\,{\bf E}^2}$ (for the positive permittivity) and $-\sqrt{{\bf k}^2+m^2-g^2\,{\bf E}^2}<\omega<\sqrt{{\bf k}^2+m^2-g^2\,{\bf E}^2}$ (for the positive permeability).
Using the equations in momentum space (\ref{kD0ktimese0})$-$(\ref{axionphi0}) and the constitutive relations (\ref{D0i}) and (\ref{H0i}), we obtain the wave equation for the components of the electric amplitude :
\begin{eqnarray}\label{Mijej}
M^{ij} \, {\bf e}_{0}^{\,\,j}=0 \; ,
\end{eqnarray}
where the matrix elements $M^{ij}$ are given by
\begin{eqnarray}\label{Mij}
M^{ij} &=& a \, \delta^{ij} + b \, k^i \, k^j
+c_{B}\, B^i \, B^j
+ c_{E} \, E^i \, E^j
+d_{B} \left({\bf B\cdot k}\right) \left(B^i \, k^j + B^j \, k^i\right)
\nonumber \\
&&
\hspace{-1cm}
+ \, d_{E} \left({\bf E\cdot k}\right)\left(E^i \, k^j + E^j \, k^i\right)
- \, d_{B} \, \omega \left[ E^i \left( {\bf B} \times {\bf k} \right)^j + E^{j}\left( {\bf B} \times {\bf k} \right)^i \right]
\nonumber \\
&&
\hspace{-1cm}
+ \, d_{E} \, \omega \left[ \, B^i \left( {\bf E} \times {\bf k} \right)^j + B^{j} \left( {\bf E} \times {\bf k} \right)^i \, \right]
\; ,
\end{eqnarray}
whose the coefficients $a$, $b$, $c_{B}$ and $c_{E}$
are defined by
\begin{subequations}
\begin{eqnarray}
a \!&=&\! \omega^2-{\bf k}^2+d_{B} \left({\bf k} \times {\bf B}\right)^2+d_{E} \left({\bf k} \times {\bf E}\right)^2
\; , \;
\\
b \!&=&\! 1-d_{B} \, {\bf B}^2-d_{E} \, {\bf E}^2
\; ,
\\
c_{B} \!&=&\! d_{E} \, \omega^2 - d_{B} \, {\bf k}^2
\; , \;
c_{E} = d_{B} \, \omega^2- d_{E} \, {\bf k}^2
\; .
\end{eqnarray}
\end{subequations}
The non-trivial solution of eq. (\ref{Mijej}) implies that the determinant of the matrix $M^{ij}$ is null.
It is difficult to analyze the situation with both ${\bf E} \neq {\bf 0}$ and ${\bf B} \neq {\bf 0}$.
The frequency solutions are feasible for the cases with ${\bf E}= {\bf 0}$ or ${\bf B}= {\bf 0}$. In what follows,
we investigate these two particular cases separately.
\subsection{The magnetic background case}
\label{subsec3A}
Let us consider ${\bf E} = {\bf 0}$ in the matrix elements (\ref{Mij}) :
\begin{eqnarray}\label{MijE=0}
\left. M^{ij}\right|_{{\bf E}=0} = a_{B} \, \delta^{ij} + b_{B} \, k^i \, k^j
+c_{B} \, B^i \, B^j
+d_{B}\left({\bf B\cdot k}\right) \left(B^i \, k^j + B^j \, k^i\right) \; ,
\end{eqnarray}
where $a_{B}=\omega^2 - {\bf k}^2 +d_{B} \left({\bf k} \times {\bf B}\right)^2$ and $b_{B}=1-d_{B} \, {\bf B}^2$.
The determinant of  (\ref{MijE=0}) is
\begin{eqnarray}\label{detME=0}
\left.\det(M)\right|_{{\bf E}=0}=a_{B} \left\{ \, a_B^2 + 2 \, a_B \, d_B \left( {\bf B}\cdot{\bf k} \right)^{2} +a_{B} \left( c_{B} {\bf B}^2+b_{B}{\bf k}^2 \right)+
\right.
\nonumber \\
\left.
+ \, b_{B}\,c_{B}\left( {\bf k} \times {\bf B} \right)^{2}-d_B^2 \, ({\bf B}\cdot{\bf k})^{2} \, ({\bf k}\times{\bf B})^{2}  \, \right\} \; ,
\end{eqnarray}
and imposing that $\left. \det(M) \right|_{{\bf E}=0}=0$, we obtain the first solution $a_{B}=0$, that yields
\begin{eqnarray}\label{omega1B}
\omega_{1B}({\bf k})= |{\bf k}| \, \sqrt{1-\frac{d_1}{c_1} \, ({\bf B} \times \hat{{\bf k}})^{2}} \; .
\end{eqnarray}
The other solutions come from the polynomial equation
\begin{eqnarray}\label{EqomegaB}
\omega^2\left(P_{B} \, \omega^4 + Q_{B} \, \omega^2+R_{B}\right)=0 \; ,
\end{eqnarray}
where the coefficients are defined by
\begin{subequations}
\begin{eqnarray}
P_{B} &=& 1+\frac{d_2}{c_1} \, {\bf B}^2 \; ,
\\
Q_{B} &=& -2 \, {\bf k}^2 - m^2 - \frac{d_2}{c_1} \, {\bf B}^2 \left({\bf k}^2+m^2\right)
-\frac{d_2}{c_1} \left({\bf k} \cdot {\bf B}\right)^2 - \frac{g^2}{c_1} \, {\bf B}^2 \; ,
\\
R_{B} &=& {\bf k}^2 \, ({\bf k}^2+m^2)+\frac{d_2}{c_1} \, ({\bf k}^2+m^2) \left( {\bf k}\cdot{\bf B}  \right)^2
+\frac{g^2}{c_1} \, ({\bf k}\cdot{\bf B})^{2} \; .
\end{eqnarray}
\end{subequations}
The second solution is the trivial $\omega=0$. Solving the above polynomial equation, one can show that the non-trivial solutions correspond to
\begin{subequations}
\begin{eqnarray}
\omega_{2B}^2={\bf k}^2+\frac{m^2}{2}+\frac{g^2{\bf B}^2-d_2({\bf B}\times{\bf k})^{2}}{2(c_1+d_2{\bf B}^2)}
\nonumber \\
-\sqrt{ \left[ {\bf k}^2+\frac{m^2}{2}+\frac{g^2{\bf B}^2-d_2({\bf B}\times{\bf k})^{2}}{2(c_1+d_2{\bf B}^2)} \right]^2
\!\!\!-{\bf k}^2 \, \frac{g^2({\bf B}\cdot{\hat{{\bf k}}})^{2}+({\bf k}^2+m^2)(c_{1}+d_2({\bf B}\cdot{\hat{{\bf k}}})^{2})}{c_1+d_2{\bf B}^2} }
\; , \;\;\;
\label{omega2B}
\\
\omega_{3B}^2={\bf k}^2+\frac{m^2}{2}+\frac{g^2{\bf B}^2-d_2({\bf B}\times{\bf k})^{2}}{2(c_1+d_2{\bf B}^2)}
\nonumber \\
+\sqrt{ \left[ {\bf k}^2+\frac{m^2}{2}+\frac{g^2{\bf B}^2-d_2({\bf B}\times{\bf k})^{2}}{2(c_1+d_2{\bf B}^2)} \right]^2
\!\!\!-{\bf k}^2 \, \frac{g^2({\bf B}\cdot{\hat{{\bf k}}})^{2}+({\bf k}^2+m^2)(c_{1}+d_2({\bf B}\cdot{\hat{{\bf k}}})^{2})}{c_1+d_2{\bf B}^2} }
\; . \;\;\;
\label{omega3B}
\end{eqnarray}
\end{subequations}
The equations (\ref{omega2B}) and (\ref{omega3B}) indicate that the dispersive character of the refractive index is exclusively due to the presence of the axion. The non-linearity alone does not yield dispersion, as the equation above show.
At this stage, it is interesting to analyse the approximation of the very small axion coupling constant. Using $g^2 \, |{\bf B}| \ll 1$, the previous frequencies are reduced to
\begin{eqnarray}\label{omega23B}
\omega_{2B}({\bf k}) \simeq \sqrt{{\bf k}^2+m^2}+{\cal O}(g^2)
\hspace{0.3cm} \mbox{and} \hspace{0.3cm}
\omega_{3B}({\bf k}) \simeq |{\bf k}| \, \sqrt{1-\frac{d_2 \, ({\bf B}\times\hat{{\bf k}})^{2} }{c_1+d_2{\bf B}^2} } +{\cal O}(g^2) \; .
\end{eqnarray}
In this approximation, $\omega_{2B}$ leads to the usual DR for a massive particle, while the results for $\omega_{1B}$ and $\omega_{3B}$, eqs. (\ref{omega1B}) and (\ref{omega23B}), confirm the DRs obtained in ref. \cite{MJNevesEDN} for a general non-linear ED in a uniform magnetic background.
The refractive index associated with the DRs are defined by
\begin{eqnarray}\label{niB}
n_{iB}=\frac{|{\bf k}|}{\omega_{iB}} \; (i=1,2,3) \; .
\end{eqnarray}
We point out that, for the DR in eq. (\ref{omega1B}), the refractive index only depends on the direction of the magnetic field
${\bf B}$ with the wave propagating direction ${\bf k}$. On the other hand, for the DRs in eqs. (\ref{omega2B}) and (\ref{omega3B}),
the refractive index depends on the wavelength $(\lambda=2\pi/|{\bf k}|)$ due to the presence of the axion mass $(m)$
and the coupling constant $(g)$. In the limit $m \rightarrow 0$ and $g\rightarrow 0$,
all the refractive indices do not depend on the wavelength in the non-linear EDs.
Since we have three solutions for the frequencies, each solution has a different group velocity.
For the frequency in  eq. (\ref{omega1B}), we obtain
\begin{eqnarray}\label{vgomega1B}
\left. {\bf v}_{gB} \right|_{\omega=\omega_{1B}}=\hat{{\bf k}} \; \sqrt{1-\frac{d_1}{c_1} \, (\hat{{\bf k}}\times{\bf B})^2 } \; .
\end{eqnarray}
The polynomial equation (\ref{EqomegaB}) has the correspondent group velocity:
\begin{eqnarray}\label{vgdomegadk}
{\bf v}_{gB}=\hat{{\bf k}} \, \frac{d\omega}{dk}=\frac{- \, \hat{{\bf k}}}{2\omega\left(2\omega^2P_{B}+Q_{B}\right)} \left(\frac{dR_{B}}{dk}+\omega^2 \, \frac{dQ_{B}}{dk} \right) \; ,
\end{eqnarray}
where $\omega$ is now evaluated at the DRs $\omega=\omega_{2B}$ and $\omega=\omega_{3B}$, in which $k \equiv |{\bf k}|$. Using the definitions of $P_{B}$, $Q_{B}$ and $R_{B}$, the expression (\ref{vgdomegadk}) is read below
\begin{equation}\label{vgomega2B3B}
{\bf v}_{gB}=
\frac{{\bf k}}{\omega}
\left[ \frac{2c_{1}(\omega^2-{\bf k}^2)-c_{1}m^2 -d_2({\bf B}\cdot{\bf k})^{2}+d_2\omega^2{\bf B}^2+d_{2}(\omega^2-{\bf k}^2-m^2)({\bf B}\cdot\hat{{\bf k}})^{2}-g^2({\bf B}\cdot\hat{{\bf k}})^{2} }{2c_{1}(\omega^2-{\bf k}^2)-c_{1}m^2+d_2{\bf B}^2(2\omega^2-{\bf k}^2-m^2)-d_{2}({\bf B}\cdot\hat{{\bf k}})^{2}-g^2{\bf B}^2} \right] \;.
\end{equation}
Substituting the frequencies $\omega_{2B}$ and $\omega_{3B}$ in eq. (\ref{vgomega2B3B}), the group velocities in the approximation   $g^2 \, |{\bf B}| \ll 1$ are given by :
\begin{subequations}
\begin{eqnarray}
\left. {\bf v}_{gB} \right|_{\omega = \omega_{2B}} &\simeq& \hat{{\bf k}} \; \sqrt{ 1 - \frac{d_2\,({\bf B}\times\hat{{\bf k}})^2}{c_1+d_2 \, {\bf B}^2} }
+{\cal O}(g^2) \; \, ,
\label{vgomega2B}
\\
\left. {\bf v}_{gB} \right|_{\omega = \omega_{3B}} &\simeq& \frac{{\bf k}}{\sqrt{{\bf k}^2+m^2}}+{\cal O}(g^2)  \,\, \; .
\end{eqnarray}
\end{subequations}
The results (\ref{vgomega1B}) and (\ref{vgomega2B}) show the dependence of the group velocity on the angle between the magnetic background ${\bf B}$ and the propagation direction $\hat{{\bf k}}$. It is important to highlight that the Maxwell limit recovers the known results for the group velocities (\ref{vgomega1B}) and (\ref{vgomega2B3B}), {\it i.e.}, ${\bf v}_{g} = c \, \hat{{\bf k}}$ (with $c=1$), when $d_1=d_2=0$ and $c_1=1$.


\subsection{The electric background case}
\label{subsec3B}
The electric background case is obtained with ${\bf B}=0$ in eq. (\ref{Mij}) :
\begin{eqnarray}\label{MijB=0}
\left. M^{ij} \right|_{{\bf B}=0} = a_{E} \, \delta^{ij} + b_{E} \, k^i \, k^j
+c_{E} \, E^i \, E^j
+d_{E} \left({\bf E\cdot k}\right)\left(E^i \, k^j + E^j \, k^i\right) \; ,
\end{eqnarray}
where $a_{E} = \omega^2-{\bf k}^2+d_{E} \left({\bf k} \times {\bf E}\right)^2$ and
$b_{E} = 1-d_{E} \, {\bf E}^2$.
The correspondent determinant is similar to the result (\ref{detME=0}) :
\begin{eqnarray}\label{detMB=0}
\left.\det(M)\right|_{{\bf B}=0}= a_{E} \left\{ \, a_E^2 + 2 \, a_E \, d_E \left( {\bf E}\cdot{\bf k} \right)^{2} +a_{E} \left( c_{E} {\bf E}^2+b_{E}{\bf k}^2 \right)
\right.
\nonumber \\
\left.
+ \, b_{E}\,c_{E}\left( {\bf k} \times {\bf E} \right)^{2}-d_E^2 \, ({\bf E}\cdot{\bf k})^{2} \, ({\bf k}\times{\bf E})^{2}  \, \right\} \; .
\end{eqnarray}
The null determinant in eq. (\ref{detMB=0}) implies the first condition $a_E =0$, or equivalently $\omega^2-{\bf k}^2+d_{E} \left({\bf k} \times {\bf E}\right)^2=0$, that yields the solutions $\omega_{1E}^{\pm}=\pm \, \omega_{1E}({\bf k})$ and $\omega_{2E}^{\pm}=\pm \, \omega_{2E}({\bf k})$,
where the DRs are given by
\begin{subequations}
\begin{eqnarray}
\omega_{1E}({\bf k})=\sqrt{ {\bf k}^2+\frac{m^2}{2}-\frac{d_2}{2c_1} \, ({\bf E}\times{\bf k})^{2}- \sqrt{ \left[\frac{m^2}{2}+\frac{d_2}{2c_1}\,({\bf E}\times{\bf k})^2 \right]^2+\frac{g^2}{c_1} \,({\bf E}\times{\bf k})^2 } } \; , \;\;
\label{reldispE1}
\\
\omega_{2E}({\bf k})=\sqrt{ {\bf k}^2+\frac{m^2}{2}-\frac{d_2}{2c_1} \, ({\bf E}\times{\bf k})^{2}+ \sqrt{ \left[\frac{m^2}{2}+\frac{d_2}{2c_1}\,({\bf E}\times{\bf k})^2 \right]^2+\frac{g^2}{c_1} \,({\bf E}\times{\bf k})^2 } } \; . \;\;
\label{reldispE2}
\end{eqnarray}
\end{subequations}
The second condition for eq. (\ref{detMB=0}) to be null leads to the polynomial equation
\begin{eqnarray}\label{EqomegaE}
\omega^2 \left[ \left(1+\frac{d_1}{c_1} \, {\bf E}^2 \right) \, \omega^2 -{\bf k}^2-\frac{d_1}{c_1} \, ({\bf E}\cdot{\bf k})^{2} \right] = 0 \; .
\end{eqnarray}
The first root in eq. \eqref{EqomegaE} is $\omega=0$, and the non-trivial solutions are
$\omega_{3E}^{\pm}=\pm \, \omega_{3E}({\bf k})$, with
\begin{eqnarray}\label{reldispE3}
\omega_{3E}({\bf k})=|{\bf k}| \, \sqrt{ \, 1 - \frac{d_{1}({\bf E}\times\hat{{\bf k}})^{2}}{c_1+d_1 {\bf E}^2} \, } \; .
\end{eqnarray}
Therefore, we obtain three possible DRs for the axionic non-linear ED in an electric background. However, only the frequencies $\omega_{1E}({\bf k})$ and $\omega_{2E}({\bf k})$ contain contributions of the axion field.
The refractive index in an electric background field is
\begin{eqnarray}
n_{iE}=\frac{|{\bf k}|}{\omega_{iE}} \; (i=1,2,3) \; .
\end{eqnarray}
The DR (\ref{reldispE3}) yields a refractive index that depends on the direction of
${\bf E}$ with the ${\bf k}$-wave propagation. In the case of the DRs (\ref{reldispE1})
and (\ref{reldispE2}), the refractive indices depend on the wavelength if we consider $m \neq 0$.

From the condition $a_{E}=0$, the correspondent group velocity is given by
\begin{eqnarray}\label{vgE}
{\bf v}_{gE} = \frac{{\bf k}}{\omega} \left[1+\frac{g^2}{c_1} \frac{({\bf E}\times{\bf k})^{2}}{({\bf k}^2-\omega^2+m^2)^{2}} \right]^{-1}\times
\nonumber \\
\times \, \left[1-\frac{d_2}{c_1} \, ({\bf E}\times\hat{{\bf k}})^{2}-\frac{g^2}{c_1} \frac{({\bf E}\times\hat{{\bf k}})^{2}}{{\bf k}^2-\omega^2+m^2}
+\frac{g^2}{c_1} \frac{({\bf E}\times{\bf k})^{2}}{({\bf k}^2-\omega^2+m^2)^{2}} \right] \; ,
\end{eqnarray}
where $\omega$ must be evaluated at the dispersion relations $\omega_{1E}$ and $\omega_{2E}$.
Substituting the frequencies (\ref{reldispE1}) and (\ref{reldispE2}) in eq. (\ref{vgE}), we obtain the results
\begin{subequations}
\begin{eqnarray}
\left. {\bf v}_{gE} \right|_{\omega=\omega_{1E}} &\simeq& \hat{{\bf k}} \; \sqrt{1-\frac{d_2}{c_1} \, ({\bf E}\times \hat{{\bf k}})^{2} } + {\cal O}(g^2) \; ,
\label{vgEsol1omega1E} \\
\left. {\bf v}_{gE} \right|_{\omega=\omega_{2E}} &\simeq& \frac{{\bf k}}{\sqrt{{\bf k}^2+m^2}} \left[1-\frac{d_2}{c_1} \, ({\bf E}\times \hat{{\bf k}})^{2} \right] +{\cal O}(g^2) \; .
\label{vgEsol1omega2E}
\end{eqnarray}
\end{subequations}
The third possible solution for the group velocity comes from the eq. (\ref{EqomegaE}). In this case, we obtain
the group velocity
\begin{eqnarray}\label{vgEsol2}
{\bf v}_{gE}= \frac{{\bf k}}{\omega} \left[ \, 1 - \frac{d_{1} \, ({\bf E}\times\hat{{\bf k}})^2}{c_1+d_1\,{\bf E}^2} \, \right] \; .
\end{eqnarray}
Using the dispersion relation (\ref{reldispE3})  in eq. (\ref{vgEsol2}), the correspondent group velocity reads
\begin{eqnarray}
\left. {\bf v}_{gE} \right|_{\omega=\omega_{3E}} = \hat{{\bf k}} \; \sqrt{ 1 - \frac{d_1\,({\bf E}\times\hat{{\bf k}})^2}{c_1+d_1 \, {\bf E}^2} }  \; . \label{vgEsol1omega_3E}
\end{eqnarray}
In the Maxwell limit, $d_{1}=d_{2} = 0$ and $c_{1}=1$, the group velocities (\ref{vgEsol1omega1E}) and (\ref{vgEsol1omega_3E}) reduce to the usual case when $g \rightarrow 0$ : ${\bf v}_{gE} =c \, \hat{{\bf k}}$ (with $c=1$). Still in this limit, the group velocity (\ref{vgEsol1omega2E})
is reduce to the result of a wave-particle of mass $m$. Analogously to the magnetic background case, the results obtained in this subsection also depends on the angle between the electric background and the wave propagation direction. In all the results, the dispersion relations and the group velocities depend on the coefficients $c_{1}$, $d_{1}$ and $d_{2}$, which are fixed by the non-linear ED as functions of the magnetic or electric background fields.


\section{Application to the axion-Born-Infeld model}
\label{sec4}
In this section, we apply the Born-Infeld (BI) theory as an example of non-linear electrodynamics in the model (\ref{Lmodel}).
Therefore, we can discuss the results of the previous section applied to a well-known non-linear ED in the literature.
The BI electrodynamics is described by the Lagrangian
\begin{eqnarray}\label{LBI}
{\cal L}_{BI}({\cal F}_{0},{\cal G}_{0})= \beta^2 \left[ \, 1-\sqrt{ 1-2\,\frac{{\cal F}_{0}}{\beta^2}-\frac{{\cal G}_{0}^2}{\beta ^4} } \, \, \right] \; ,
\end{eqnarray}
where $\beta$ is a scale parameter with dimension of squared energy (in natural units). It can be interpreted as a critical field in the theory, and has the same dimension of the electromagnetic field. The BI Lagrangian is CP-invariant since it depends on the ${\cal G}_{0}^2$. The limit of
$\beta \rightarrow \infty$ yields the Maxwell Lagrangian. The measurement by ATLAS of the light-by-light scattering in Pb-Pb collisions constraints a stringent lower bound on the $\beta$-parameter \cite{Ellis}, namely,
\begin{eqnarray}
\sqrt{\beta} \, \gtrsim \, 100 \, \mbox{GeV} \; ,
\end{eqnarray}
in the case of a pure quantum electrodynamics with the Born-Infeld theory associated with the Abelian $U(1)$ symmetry. Keeping this in mind, we shall consider $\sqrt{\beta} =100$ GeV as the scale of the BI theory in our future analysis. In what follows, let us examine the dispersion relations and group velocities, as well as the properties of the permittivity and permeability tensors of the axion-BI model for the cases of magnetic background field  (subsection VI.A), and posteriorly, in the presence of an electric background field  (subsection VI.B).
\subsection{The axion-BI model in a magnetic background}
\label{subsec4A}
From eq. (\ref{coefficients}), we obtain the correspondent coefficients $c_1$, $d_1$ and $d_2$ for the BI theory,
\begin{eqnarray}\label{c1d1d2BIB}
\left. c_{1}^{BI} \right|_{{\bf E}=0,{\bf B}}=\frac{\beta}{ \sqrt{\beta^2+{\bf B}^2} }
\; , \;
\left.d_{1}^{BI} \right|_{{\bf E}=0,{\bf B}} = \frac{\beta}{(\beta^2+{\bf B}^2)^{3/2}}
\; , \;
\left. d_{2}^{BI} \right|_{{\bf E}=0,{\bf B}}=\frac{1}{\beta \sqrt{\beta^2+{\bf B}^2}  } \; ,
\end{eqnarray}
with $c_{2}=d_{3}=0$. All these coefficients are positive and the axion-BI model reduces to the usual Maxwell theory coupled to the axion in the limit $\beta \rightarrow \infty$, {\it i. e.}, $\lim_{\beta\rightarrow \infty} \left. c_{1}^{BI} \right|_{{\bf E}=0,{\bf B}}=1$ and $\lim_{\beta\rightarrow \infty} \left. d_{1}^{BI} \right|_{{\bf E}=0,{\bf B}}= \lim_{\beta\rightarrow \infty} \left. d_{2}^{BI} \right|_{{\bf E}=0,{\bf B}} =0$ . Substituting these coefficients in the results (\ref{omega1B}), (\ref{omega2B}) and (\ref{omega3B}), we obtain the dispersion relations in a uniform and constant magnetic field ${\bf B}$ :
\begin{subequations}
\begin{eqnarray}
\omega_{1B}^{(BI)}({\bf k}) &=& |{\bf k}| \, \sqrt{1- \frac{({\bf B} \times \hat{{\bf k}})^{2}}{\beta^2+{\bf B}^2}} \; .
\label{omega1BBI}
\\
\left[\omega_{2B}^{(BI)}({\bf k})\right]^2 &=& {\bf k}^2+\frac{m^2}{2}-\frac{({\bf B}\times{\bf k})^{2}}{2 \left(\beta^2+{\bf B}^2\right)}
+\frac{g^2 \, {\bf B}^2}{2\sqrt{1+{\bf B}^{2}/\beta^2}}
\label{omega2BBI}
\nonumber \\
&&
- \left\{\left[{\bf k}^2+\frac{m^2}{2}-\frac{({\bf B}\times{\bf k})^{2}}{2 \left(\beta^2+{\bf B}^2\right)}
+\frac{g^2\,{\bf B}^2}{2\sqrt{1+{\bf B}^{2}/\beta^2}}\right]^2
\right.
\nonumber \\
&&
\left.
-{\bf k}^2 \left({\bf k}^2+m^2\right) \left[1-\frac{({\bf B}\times\hat{{\bf k}})^2}{\beta^2+{\bf B}^2}\right]\!-\frac{g^2\,({\bf B}\cdot{\bf k})^2}{\sqrt{1+{\bf B}^2/\beta^2}} \right\}^{1/2} ,
\\
\left[\omega_{3B}^{(BI)}({\bf k})\right]^{2} &=&
{\bf k}^2+\frac{m^2}{2}-\frac{({\bf B}\times{\bf k})^{2}}{2 \left(\beta^2+{\bf B}^2\right)}
+\frac{g^2\,{\bf B}^2}{2\sqrt{1+{\bf B}^{2}/\beta^2}}
\nonumber \\
&&
+ \left\{\left[{\bf k}^2+\frac{m^2}{2}-\frac{({\bf B}\times{\bf k})^{2}}{2 \left(\beta^2+{\bf B}^2\right)}
+\frac{g^2\,{\bf B}^2}{2\sqrt{1+{\bf B}^{2}/\beta^2}}\right]^2
\right.
\nonumber \\
&&
\left.
-{\bf k}^2 \left({\bf k}^2+m^2\right) \left[1-\frac{({\bf B}\times\hat{{\bf k}})^2}{\beta^2+{\bf B}^2}\right]\!
-\frac{g^2\,({\bf B}\cdot{\bf k})^2}{\sqrt{1+{\bf B}^2/\beta^2}} \right\}^{1/2} \; .
\label{omega3BBI}
\end{eqnarray}
\end{subequations}
At this stage, it is interesting to consider particular limits. For instance,
under an intense magnetic background field, {\it i. e.}, $|{\bf B}| \gg \beta$, the dispersion relations are reduced to
$\omega_{1B}^{(BI)}({\bf k}) = \omega_{2B}^{(BI)}({\bf k}) \simeq |{\bf k}\cdot\hat{{\bf B}}|$ and
$\omega_{3B}^{(BI)}({\bf k}) \simeq \sqrt{{\bf k}^2+m^2}$, in which we have also considered
$g^2\beta \ll 1$ in the frequencies $\omega_{2B}^{(BI)}$ and $\omega_{3B}^{(BI)}$. In the strong magnetic field regime, the DRs
$\omega_{1B}^{(BI)}$ and $\omega_{2B}^{(BI)}$ are not dependent on the magnetic field magnitude, and depend only on the angle between the wave
vector ${\bf k}$ and the direction of the magnetic background field. On the other hand, when $\beta\rightarrow \infty$, we recover the photon DR in the frequency $\omega_{1B}^{(BI)}$, while $\omega_{2B}^{(BI)}$ and $\omega_{3B}^{(BI)}$ reduce to the expressions :
\begin{subequations}
\begin{eqnarray}
\lim_{\beta\rightarrow\infty}\omega_{2B}^{(BI)}({\bf k}) &=& \sqrt{ \, {\bf k}^2
+\frac{m^2+g^2\,{\bf B}^2}{2}
-\sqrt{\left(\frac{m^2+g^2\,{\bf B}^2}{2}\right)^2+g^2\left({\bf B}\times{\bf k}\right)^2} } \; ,
\label{omega2Bbetainfinity}
\\
\lim_{\beta\rightarrow\infty}\omega_{3B}^{(BI)}({\bf k}) &=& \sqrt{ \, {\bf k}^2
+\frac{m^2+g^2\,{\bf B}^2}{2}
+\sqrt{\left(\frac{m^2+g^2\,{\bf B}^2}{2}\right)^2+g^2\left({\bf B}\times{\bf k}\right)^2} } \; .
\label{omega3Bbetainfinity}
\end{eqnarray}
\end{subequations}
The dependence of the DRs with the angle between ${\bf B}$ and ${\bf k}$ remains in eqs. (\ref{omega2Bbetainfinity}) and (\ref{omega3Bbetainfinity}).
In the regime $g^2|{\bf B}| \ll 1$, the previous frequencies lead to
\begin{subequations}
\begin{eqnarray}
\omega_{2B}({\bf k}) &\simeq& |{\bf k}| \left[1-\frac{g^2}{2m^2} \left({\bf B}\times\hat{{\bf k}}\right)^2 \right] \; ,
\label{omega2Bbetainfinitygpq}
\\
\omega_{3B}({\bf k}) &\simeq& \sqrt{ {\bf k}^2+m^2} + \frac{g^2 \, {\bf B}^2}{2\sqrt{{\bf k}^2+m^2}}
+ \frac{g^2}{2m^2} \frac{\left( {\bf B}\times{\bf k} \right)^2}{\sqrt{{\bf k}^2+m^2}} \; .
\label{omega3Bbetainfinitygpq}
\end{eqnarray}
\end{subequations}
Using the results (\ref{vgomega1B}) and (\ref{vgomega2B}), the group velocity for the axion-BI model is
\begin{eqnarray}
\left. {\bf v}_{gBI} \right|_{\omega_{1B}^{(BI)}}\!=
\left. {\bf v}_{gBI} \right|_{\omega_{2B}^{(BI)}}\!=\hat{{\bf k}} \,\, \sqrt{1-\frac{({\bf B} \times \hat{{\bf k}})^{2}}{\beta^2+{\bf B}^2}} \; .
\end{eqnarray}
As expected, the $\beta \rightarrow \infty$ limit recovers the result of usual electrodynamics group velocity ${\bf v}_{g}=\hat{{\bf k}}$, in natural units. Under a strong magnetic field, the group velocities also depend on the angle of $\hat{{\bf k}}$ with the $\hat{{\bf B}}$-direction :
\begin{eqnarray}
\left. {\bf v}_{gBI} \right|_{\omega_{1B}^{(BI)}}\!=
\left. {\bf v}_{gBI} \right|_{\omega_{2B}^{(BI)}}\!\simeq \hat{{\bf k}} \, |\hat{{\bf k}}\cdot\hat{{\bf B}}|=\hat{{\bf k}} \, |\cos\theta| \; .
\end{eqnarray}
Substituting the coefficients \eqref{c1d1d2BIB} in the eigenvalues of the electric permittivity matrix, we obtain
\begin{subequations}
\begin{eqnarray}
\left. \lambda_{1\varepsilon}^{(BI)} \right|_{{\bf E}={\bf 0}}
\!&=&\! \left. \lambda_{2\varepsilon}^{(BI)} \right|_{{\bf E}={\bf 0}} = \frac{\beta}{\sqrt{\beta^2 + {\bf B}^2}} \; ,
\label{lambda1eBI}
\\
\left. \lambda_{3\varepsilon}^{(BI)} \right|_{{\bf E}={\bf 0}} \!&=&\! \sqrt{1+\frac{{\bf B}^2}{\beta^2}}+ \frac{g^2 \, {\bf B}^2}{ {\bf k}^2 +m^2 - \omega^2} \, .
\label{lambda3eBI}
\end{eqnarray}
\end{subequations}
Similarly, the eigenvalues of the magnetic permeability are
\begin{eqnarray}\label{lambdamuBI}
\left. \lambda_{1\mu}^{(BI)} \right|_{{\bf E}={\bf 0}}
= \left. \lambda_{2\mu}^{(BI)} \right|_{{\bf E}={\bf 0}} = \sqrt{1 + \frac{{\bf B}^2}{\beta^2}}
\hspace{0.4cm} , \hspace{0.4cm}
\left. \lambda_{3\mu}^{(BI)} \right|_{{\bf E}={\bf 0}} = \left(1 + \frac{{\bf B}^2}{\beta^2} \right)^{3/2} \; .
\end{eqnarray}
With these expressions, we conclude that the eigenvalues of eqs. (\ref{lambda1eBI}) and (\ref{lambdamuBI}) are positive, while that in eq. (\ref{lambda3eBI}) is positive if the $\omega$-frequency satisfies the inequality :
\begin{eqnarray}\label{condomegaBIB}
- \, \sqrt{ {\bf k}^2+m^2+\frac{g^2\,{\bf B}^2}{\sqrt{1+{\bf B}^2/\beta^2}} } <\omega< \sqrt{ {\bf k}^2+m^2+\frac{g^2\,{\bf B}^2}{\sqrt{1+{\bf B}^2/\beta^2}} } \; .
\end{eqnarray}
In the limit $\beta \rightarrow \infty$, the results (\ref{lambda1eBI}) and (\ref{lambdamuBI}) go to one, but in  eq.  (\ref{lambda3eBI}), we obtain a contribution of the axion coupling with the Maxwell ED :
\begin{eqnarray}
\lim_{\beta \rightarrow \infty} \left. \lambda_{3\varepsilon}^{(BI)} \right|_{{\bf E}={\bf 0}} = 1 + \frac{g^2 \, {\bf B}^2}{ {\bf k}^2 +m^2 - \omega^2} \, .
\end{eqnarray}
\subsection{The axion-BI model in an electric background}
\label{subsec4B}
In this case, the coefficients $c_{1}$, $d_{1}$ and $d_{2}$ at ${\bf B}=0$ are given by
\begin{eqnarray}\label{c1d1d2BIE}
\left. c_{1}^{BI} \right|_{{\bf E},{\bf B}=0}=\frac{\beta}{ \sqrt{\beta^2-{\bf E}^2} }
\; , \;
\left.d_{1}^{BI} \right|_{{\bf E},{\bf B}=0} = \frac{\beta}{(\beta^2-{\bf E}^2)^{3/2}}
\; , \;
\left. d_{2}^{BI} \right|_{{\bf E},{\bf B}=0}=\frac{1}{\beta \sqrt{\beta^2-{\bf E}^2}  } \; ,
\end{eqnarray}
and $c_{2}=d_{3}=0$, in which the magnitude of the electric background must satisfy the condition $ \beta > |{\bf E}|$ for
these coefficients to be real. When $|{\bf E}| > \beta$, the coefficients are complex and can bring interesting consequences for
the DRs and the wave propagation. Phenomenologically, the constraint of $|{\bf E}| > \beta$ is not usual since that,
in general, electric fields with strong magnitude of $|{\bf E}| > 10^2 \, \mbox{GeV}^2$ are not suitable. Using these coefficients in the dispersion relations (\ref{reldispE1}), (\ref{reldispE2}) and (\ref{reldispE3}), we arrive at the following frequencies for the axion-BI model in a uniform and constant electric background :
\begin{subequations}
\begin{eqnarray}
\omega_{1E}^{(BI)}({\bf k}) \!&=&\! \sqrt{ {\bf k}^2+\frac{m^2}{2}-\frac{({\bf E}\times{\bf k})^{2}}{2 \beta ^2}- \sqrt{\left[\frac{m^2}{2}+\frac{({\bf E}\times{\bf k})^{2}}{2\beta^2}\right]^2+g^2 ({\bf E}\times{\bf k})^{2} \sqrt{1-\frac{{\bf E}^2}{\beta ^2}}}} ,
\hspace{0.5cm}
\label{omega1EBI}
\\
\omega_{2E}^{(BI)}({\bf k}) \!&=&\! \sqrt{ {\bf k}^2+\frac{m^2}{2}-\frac{({\bf E}\times{\bf k})^{2}}{2 \beta ^2}+ \sqrt{\left[\frac{m^2}{2}+\frac{({\bf E}\times{\bf k})^{2}}{2\beta^2}\right]^2+g^2 ({\bf E}\times{\bf k})^{2} \sqrt{1-\frac{{\bf E}^2}{\beta ^2}}}} ,
\hspace{0.5cm}
\label{omega2EBI}
\\
\omega_{3E}^{(BI)}({\bf k}) &=& |{\bf k}| \, \sqrt{1- \frac{({\bf E} \times \hat{{\bf k}})^{2}}{\beta^2}} \; ,
\label{omega3EBI}
\end{eqnarray}
\end{subequations}
where $\beta > |{\bf E} \times \hat{{\bf k}}|$ in the frequency (\ref{omega3EBI}). Notice that the Maxwell limit yields the photon DR in eq. (\ref{omega3EBI}). Moreover, the solutions (\ref{omega1EBI}) and (\ref{omega2EBI}) are, respectively, reduced to
\begin{subequations}
\begin{eqnarray}
\lim_{\beta\rightarrow\infty} \omega_{1E}^{(BI)}({\bf k}) &=& \sqrt{ {\bf k}^2+\frac{m^2}{2}- \sqrt{ \frac{m^4}{4}+g^2 \left({\bf E}\times{\bf k}\right)^{2}}} \; ,
\hspace{0.5cm}
\label{limomega1EBI}
\\
\lim_{\beta\rightarrow\infty} \omega_{2E}^{(BI)}({\bf k}) &=& \sqrt{ {\bf k}^2+\frac{m^2}{2}+ \sqrt{\frac{m^4}{4}+g^2 \left({\bf E}\times{\bf k}\right)^{2} }} \; .
\hspace{0.5cm}
\label{limomega2EBI}
\end{eqnarray}
\end{subequations}
For a weak electric field, the previous DRs assume the form
\begin{subequations}
\begin{eqnarray}
\omega_{1E}({\bf k}) &\simeq& |{\bf k}| \left[ \, 1 - \frac{g^2}{2m^2}\left({\bf E}\times\hat{{\bf k}}\right)^{2} \, \right] \; ,
\hspace{0.5cm}
\label{limomega1EBIgpeq}
\\
\omega_{2E}({\bf k}) &\simeq& \sqrt{ {\bf k}^2+m^2 } + \frac{g^2}{2m^2} \frac{\left({\bf E}\times{\bf k}\right)^{2}}{\sqrt{{\bf k}^2+m^2}} \; .
\hspace{0.5cm}
\label{limomega2EBIgpeq}
\end{eqnarray}
\end{subequations}

%
%
The group velocities associated with these dispersion relations can be read below :
\begin{subequations}
\begin{eqnarray}
\left. {\bf v}_{g} \right|_{\omega_{1E}^{(BI)}} &=& \left. {\bf v}_{g} \right|_{\omega_{3E}^{(BI)}} =  \hat{{\bf k}} \,\, \sqrt{1-\frac{({\bf E} \times \hat{{\bf k}})^{2}}{\beta^2}} \; ,
\label{vgEomega13BI}
\\
\left. {\bf v}_{g} \right|_{\omega_{2E}^{(BI)}} &=& \frac{{\bf k}}{\sqrt{{\bf k}^2+m^2}}\left[ 1-\frac{({\bf E} \times \hat{{\bf k}})^{2}}{\beta^2} \right] \; ,
\label{vgomega2EBI}
\end{eqnarray}
\end{subequations}
where the condition $\beta > |{\bf E} \times \hat{{\bf k}}|$ constraints the velocity (\ref{vgomega2EBI}) in the same direction of the wave propagation.
Analogously to the magnetic background case, we insert the expressions (\ref{c1d1d2BIE}) in the eigenvalues of the electric permittivity and magnetic permeability matrices. Thereby, we obtain for the electric permittivity :
\begin{equation}\label{lambda1eBIE}
\left. \lambda_{1\varepsilon}^{(BI)} \right|_{{\bf B}={\bf 0}} = \left. \lambda_{2\varepsilon}^{(BI)} \right|_{{\bf B}={\bf 0}}
= \frac{\beta}{\sqrt{\beta^2 - {\bf E}^2}}
\hspace{0.4cm} , \hspace{0.4cm}
\left. \lambda_{3\varepsilon}^{(BI)} \right|_{{\bf B}={\bf 0}} = \frac{\beta^3}{({\beta^2 - {\bf E}^2)^{3/2}}} \; .
\end{equation}
In addition, for the magnetic permeability, we arrive at the following eigenvalues
\begin{subequations}
\begin{eqnarray}
\left. \lambda_{1\mu}^{(BI)} \right|_{{\bf B}={\bf 0}} &=& \left. \lambda_{2\mu}^{(BI)} \right|_{{\bf B}={\bf 0}}
= \sqrt{1 - \frac{{\bf E}^2}{\beta^2} } \; ,
\label{lambda1muBIE}
\\
\left. \lambda_{3\mu}^{(BI)} \right|_{{\bf B}={\bf 0}} &=& \left[ \; \sqrt{ 1- \frac{{\bf E}^2}{\beta^2} } - \frac{g^2 \, {\bf E}^2}{ {\bf k}^2 +m^2 - \omega^2} \; \right]^{-1} \; .
\label{lambda3muBIE}
\end{eqnarray}
\end{subequations}
Observe that the eigenvalues (\ref{lambda1eBIE}) and (\ref{lambda1muBIE}) are real with the condition $\beta > |{\bf E}|$.
The eigenvalue (\ref{lambda3muBIE}) is positive if the $\omega$-frequency satisfies the condition :
\begin{eqnarray}\label{condomegaBIE}
-\sqrt{ \, {\bf k}^2+m^2-\frac{g^2\,{\bf E}^2}{\sqrt{1-{\bf E}^2/\beta^2}} \, }<\omega<\sqrt{ \, {\bf k}^2+m^2-\frac{g^2\,{\bf E}^2}{\sqrt{1-{\bf E}^2/\beta^2}} \, } \; .
\end{eqnarray}
\section{ The birefringence in the non-linear ED axion model}
\label{sec5}
The vacuum birefringence is one of the phenomena present in some non-linear EDs. We back to the wave equation (\ref{Mijej})
to investigate the birefringence in a uniform and constant magnetic background, and also for the electric background  case
in the linearized axion-BI model.
\subsection{Birefringence in the axion-BI model with a magnetic background field}
\label{subsec5A}
The birefringence analysis requires that we impose some considerations on the wave propagation. We assume the propagation direction  ${\bf k}=k\,\hat{{\bf x}}$ and the magnetic background  ${\bf B}=B \, \hat{{\bf z}}$. In the first situation, we consider the electric wave amplitude parallel to ${\bf B}$, with ${\bf e}_{0}=e_{03} \, \hat{{\bf z}}$. In this case, the wave equation (\ref{Mijej}) (with ${\bf E}={\bf 0}$) yields the relation $\mu_{22}(k,\omega)\,\varepsilon_{33}(k,\omega)\,\omega^2=k^2$, where the parallel refractive index is defined by
\begin{eqnarray}\label{nparalelB}
n_{\parallel}^{(B)}(k,\omega)=\sqrt{ \mu_{22}(k,\omega)\,\varepsilon_{33}(k,\omega) } = \sqrt{ \, 1+ \frac{d_2}{c_1}\,B^2+\frac{g^2}{c_1}\frac{B^2}{k^2-\omega^2+m^2} \, } \; .
\end{eqnarray}
The second situation is when the electric wave amplitude is perpendicular to the magnetic background field,
with ${\bf e}_{0}=e_{02} \, \hat{{\bf y}}$. In this case, the wave equation leads to the relation $\mu_{33}(k,\omega)\,\varepsilon_{22}(k,\omega)\,\omega^2=k^2$, in which the perpendicular refractive index is
\begin{eqnarray}\label{nperpendicularB}
n_{\perp}^{(B)} (k,\omega) =\sqrt{ \mu_{33}(k,\omega)\,\varepsilon_{22}(k,\omega) } = \left[ \, 1 - \frac{d_1}{c_1}\, B^2 \, \right]^{-1/2} \; .
\end{eqnarray}
Using the coefficients of the axion-BI model in eq. (\ref{c1d1d2BIB}), the difference between these two refractive indices, $\Delta n_{BI}^{(B)}(k,\omega)= n_{\parallel}^{(B)}(k,\omega) - n_{\perp}^{(B)} (k,\omega)$, is given by
\begin{eqnarray}\label{DeltanB}
\Delta n_{BI}^{(B)}(k,\omega)
=
\sqrt{ 1+\frac{B^2}{\beta^2}+ \frac{g^2 \, B^2}{k^2-\omega^2+m^2} \, \sqrt{ 1+ \frac{B^2}{\beta^2} } } -
\sqrt{ 1+ \frac{B^2}{\beta^2} } \; .
\end{eqnarray}
From eq. (\ref{nparalelB}), we expect that the variation of the refractive index depends on the $\omega$-frequency.
Thereby, we can have the birefringence phenomena associated with the three DRs from (\ref{omega1BBI}),
(\ref{omega2BBI}) and (\ref{omega3BBI}), respectively. The limit $g \rightarrow 0$ recovers the well-known result
in which the pure BI theory does not exhibit birefringence, {\it i. e.}, $\lim_{g\rightarrow 0}\Delta n_{BI}^{(B)}(k,\omega)=0$.
Substituting the dispersion relations (\ref{omega1BBI}), (\ref{omega2BBI}) and (\ref{omega3BBI}) in eq. (\ref{DeltanB}), we obtain
the differences of the refractive indices :
\begin{subequations}
\begin{eqnarray}
\left. \Delta n_{BI}^{(B)}(k) \right|_{\omega_{1B}^{(BI)}} &=& \left. \Delta n_{BI}^{(B)}(k) \right|_{\omega_{2B}^{(BI)}} \simeq \frac{g^2B}{2} \frac{ B \, (B^2+\beta^2)}{m^2 \, (B^2+\beta^2)+B^2 \, k^2} \; ,
\label{DeltanBgoomega1B}
\\
\left. \Delta n_{BI}^{(B)}(k) \right|_{\omega_{3B}^{(BI)}} &\simeq& \frac{k}{\sqrt{k^2+m^2}}
-\sqrt{1+\frac{B^2}{\beta^2}} \; ,
\label{DeltanBgoomega3B}
\end{eqnarray}
\end{subequations}
where we have used the approximation $g^2\,B \ll 1$. Notice that in eq. (\ref{DeltanBgoomega1B}), a very small (residual)
birefringence remains in the model with $g^{2}$-dependence. The result (\ref{DeltanBgoomega3B})
recovers the usual Maxwell ED when $\beta \rightarrow \infty$, and if we consider a very small mass for the
axion-particle $m\simeq0$.
The PVLAS-FE experiment presented the following result for the vacuum magnetic birefringence in ALPs \cite{25years} :
\begin{eqnarray}\label{DeltanPVLAS}
\frac{\Delta n^{\mbox{PVLAS}-\mbox{FE}}}{B^2}=(+19 \pm 27) \times 10^{-24} \, \mbox{T}^{-2} \; .
\end{eqnarray}
In this situation, we have that $B=2.5$ T and the wavelength of $\lambda=1064$ nm (or equivalently, $k=0.185$ eV).
Let us consider the axion mass at $m=1$ meV and BI parameter of $\sqrt{\beta}=100$ GeV. Using the results in eqs.
(\ref{DeltanBgoomega1B}) and (\ref{DeltanPVLAS}), we estimate the axion coupling constant as
\begin{eqnarray}\label{gresult}
g \simeq 9.065 \times 10^{-9} \, \mbox{GeV}^{-1} \; ,
\end{eqnarray}
which is consistent with the upper bound $g < 6.4 \times 10^{-8} \, \mbox{GeV}^{-1} \, (95 \% \, \mbox{C.L.})$
reported in this experiment.

\subsection{Birefringence in the axion-BI model with an electric background field}
\label{subsec5B}
The case with an electric background is similar to the previous subsection. We consider the same wave
propagation direction, with the electric background ${\bf E}=E \, \hat{{\bf z}}$. In the first situation
in which the electric wave amplitude ${\bf e}_{0}$ is parallel to ${\bf E}$, the correspondent refractive index is
\begin{eqnarray}
n_{\parallel}^{(E)}=\sqrt{ 1+ \frac{d_{1}}{c_{1}} \, E^2 } \; .
\end{eqnarray}
On the other hand, when ${\bf e}_{0}$ is perpendicular to ${\bf E}$, the refractive index leads to
\begin{eqnarray}
n_{\perp}^{(E)}(k,\omega)= \left[ 1-\frac{d_2}{c_1}\,E^2-\frac{g^2}{c_1} \, \frac{E^2}{k^2-\omega^2+m^2} \right]^{-1/2} \; .
\end{eqnarray}
Thereby, the difference between the parallel and perpendicular refractive indices in the axion-BI model is
\begin{eqnarray}\label{DeltanE}
\Delta n_{BI}^{(E)}(k,\omega)=\left(1-\frac{E^2}{\beta^2}\right)^{-1/2} \!\!\! - \left[ \, 1-\frac{E^2}{\beta^2}- \frac{g^2 \, E^2 }{k^2-\omega^2+m^2} \, \sqrt{ 1-\frac{E^2}{\beta^2} } \, \right]^{-1/2}  \;.
\end{eqnarray}
The limit of the pure BI electrodynamics $(g \rightarrow 0)$ recovers the result of null birefringence in eq. (\ref{DeltanE}).
Furthermore, note that the difference (\ref{DeltanE}) depends on the $\omega$-function, that is function of the electric field and $\beta$-parameter. Substituting the dispersion relations (\ref{omega1EBI}), (\ref{omega2EBI}) and (\ref{omega3EBI}) in eq. (\ref{DeltanE}), we obtain
\begin{subequations}
\begin{eqnarray}
\left. \Delta n_{BI}^{(E)}(k) \right|_{\omega_{1E}^{(BI)}} & = &
\left. \Delta n_{BI}^{(E)}(k) \right|_{\omega_{3E}^{(BI)}} \simeq
-\frac{g^2\,E}{2} \frac{E \, \beta^{2}}{\left(1-E^2/\beta^2\right)(m^2\beta^2+E^2k^2)} \; ,
\label{DeltanEgoomega1E}
\\
\left. \Delta n_{BI}^{(E)}(k) \right|_{\omega_{2E}^{(BI)}} &\simeq& \left( \, 1 -\frac{E^2}{\beta^2} \, \right)^{-1/2} \!\!\! -\frac{k}{\sqrt{k^2+m^2}}
\label{DeltanEgoomega2E}
\; , \;\;\;
\end{eqnarray}
\end{subequations}
in which the approximation for a weak electric field, $g^2E \ll 1$, is applied. The result (\ref{DeltanEgoomega1E}) shows a very small effect of the birefringence with the $g^2$ dependence. In eq. (\ref{DeltanEgoomega2E}), the birefringence is null when $\beta \rightarrow \infty$ and
the axion mass is approximately zero. In the approximation $\beta \gg Ek/m$, the result (\ref{DeltanEgoomega1E}) yields the electric birefringence
\begin{eqnarray}
\frac{|\Delta n_{BI}^{(E)}|}{E^2}\simeq \frac{g^2}{2m^2}
\; .
\end{eqnarray}
Therefore,  this result can be used to constraint the parameter space $(m,g)$
through the optical Kerr effect.


%
\section{Concluding Comments}
\label{sec6}
In this paper, we investigate the properties of the electromagnetic (EM)
wave propagation in a general non-linear electrodynamics coupled to the axion field.
The non-linear sector is expanded up to the second order at the propagation fields in a uniform and constant EM background field. We obtain the correspondent field equations for the propagating EM and axion fields. Next, we investigate the dispersion relations
and group velocities in the presence of magnetic and electric background fields, separately. The refractive indices associated with these dispersion relations depend on the wavelength due to contributions involving the axion mass $(m)$ and coupling constant $(g)$ in the magnetic background field.
For the case of the electric background, the dependence on the wavelength occurs when $m \neq 0$. The permittivity and permeability tensors are calculated as function of the EM background and wave propagation frequencies. We study the conditions for these tensors to be positive
through the eigenvalues of the permittivity and permeability matrices.
We apply all the results for the Born-Infeld (BI) electrodynamics coupled to the axion field. Consequently, we obtain the
wave propagation properties for the BI-axion model in magnetic and electric background fields. The results
of the usual Maxwell electrodynamics are recovered when the BI parameter is very large, and the coupling with the axion goes to zero. The magnetic permeability is manifestly positive in the BI-axion model with a magnetic background, while
one of the solutions for the electric permittivity to be positive imposes the condition (\ref{condomegaBIB}) on the $\omega$-frequency.
Otherwise, if the $\omega$-frequency satisfies the constraints
\begin{eqnarray}\label{condomegaBIBmeta}
\omega < - \, \sqrt{ {\bf k}^2+m^2+\frac{g^2\,{\bf B}^2}{\sqrt{1+{\bf B}^2/\beta^2}} }
\hspace{0.5cm} \mbox{and} \hspace{0.5cm}
\omega > \sqrt{ {\bf k}^2+m^2+\frac{g^2\,{\bf B}^2}{\sqrt{1+{\bf B}^2/\beta^2}} } \; ,
\end{eqnarray}
the medium can behave as a metamaterial. Similarly, in the axion-BI model with an
electric background, the electric permittivity is positive if $\beta > |{\bf E}|$,
and one of the solutions for the magnetic permeability to be positive constrains the
$\omega$-frequency  in eq. (\ref{condomegaBIE}).
To conclude, we have investigated the birefringence phenomena in the BI-axion model for the situations with
magnetic and electric background fields, separately. We obtain the variation of the refractive indices (parallel and perpendicular to the electric propagating amplitude) as function of the $k$-wave number, $\omega$-frequency, $\beta$-parameter and background fields in eqs. (\ref{DeltanB}) and  (\ref{DeltanE}). Therefore, since we have three solutions for the $\omega$-frequencies, the variation of the
refractive index has also three possible solutions for the birefringence. In both the cases
with magnetic or electric background fields, in the weak field regime, the variation of the refractive index is residual in $g^2$, see eqs. (\ref{DeltanBgoomega1B}) and (\ref{DeltanEgoomega1E}). The third solution points out for the birefringence as a function of the $\beta$-parameter, axion mass and background fields, as described in eqs. (\ref{DeltanBgoomega3B}) and (\ref{DeltanEgoomega2E}).
As we expect from usual electrodynamics, the absence of birefringence in these results is attained in the
$\beta \rightarrow \infty$ limit, and for the axion mass approaching zero.
In the case of the birefringence with magnetic background field, we use the results of the PVLAS-FE experiment,
with $\sqrt{\beta}=100$ GeV, $\lambda=1064$ nm and $m=1$ meV, to obtain the axion coupling constant
$g = 9.065 \times 10^{-9} \, \mbox{GeV}^{-1}$. This result agrees with the upper bound $g < 6.4 \times 10^{-8} \, \mbox{GeV}^{-1} \, (95 \% \, \mbox{C.L.})$ in the PVLAS-FE experiment. The birefringence in the electric background case gives the result $g^2/(2m^{2})$ in the regime
of weak electric field. This result must be interesting in connection with the optical Kerr effect to constraint the parameter space $(m,g)$ for the axion-like particle.

\appendix
\section{The energy-momentum tensor of the axionic non-linear electrodynamics model}
\label{appendix}

In this Appendix, we address the recent discussion about the definition of Poynting vector in axionic electrodynamics. According to ref. \cite{Tobar_PRD}, one could define two possible Poynting vectors in terms of the constitutive relations and electromagnetic fields, namely, $ {\bf S}_{DB} \sim {\bf D} \times {\bf B}$ or $ {\bf S}_{EH} \sim {\bf E} \times {\bf H}$, and the authors claim that the choice leads to different phenomenological results. This issue has also been discussed in refs. \cite{Zhou,Patkos}, where the authors point out the relevance of considering terms of the order $O(g^2)$, to analyze the corresponding conservation law. Here, we follow our own path to tackle the question: we avoid to start off from any definition; instead, we work exclusively with the field equations to naturally identify the expressions for the Poynting vector and the momentum density transported by the waves.

%
%

%
In what follows, we work out the energy-momentum tensor for the non-linear ED model coupled to the axion field in  an electromagnetic background, as described by the Lagrangian \eqref{L4phitil}. For this purpose, we contract the eq. (\ref{EqGmunu}) with $f^{\nu\alpha}$ and using the Bianchi identity, we arrive at
\begin{eqnarray} \label{eq_T1}
& \partial^\mu & \left[ c_{1} \, f_{\mu\nu} f^{\nu\alpha}
-\frac{1}{2} \, Q_{B\mu\nu\kappa\lambda} \, f^{\kappa\lambda}  f^{\nu\alpha}
+  g \, \widetilde{\phi} \, \widetilde{F}_{B\mu\nu} \, f^{\nu \alpha} \right.
-\delta_{\mu}^{\;\;\, \alpha} \left( -\frac{1}{4} \, c_{1} \, f_{\rho\sigma}^{\, 2}
+\frac{1}{8} \, Q_{B\rho\sigma\kappa\lambda} \, f^{\rho\sigma} f^{\kappa\lambda} \right.
 \nonumber \\
&-& \left. \left. \frac{1}{2} \, g \, \widetilde{\phi} \, \widetilde{F}_{B \, \rho\sigma} \, f^{\rho\sigma} \right) \right] =
J_\nu \, f^{\nu \alpha}
+ \frac{1}{4} \left(\partial^{\alpha} c_{1} \right) f_{\mu\nu}^{\, 2}
+ \frac{1}{4} \left(\partial^{\alpha} c_{2} \right) \widetilde{f}_{\mu\nu} f^{\mu\nu}
- \frac{1}{8} \, \left(\partial^{\alpha} Q_{B\mu\nu\kappa\lambda}\right) f^{\mu\nu} f^{\kappa\lambda}
 \nonumber \\
&+& \frac{1}{2} \, g \, ( \partial^\alpha \widetilde{\phi} ) \, \widetilde{F}_{B \, \mu \nu} \, f^{\mu \nu}
+ \frac{1}{2} \, g \, \widetilde{\phi} \left( \partial^\alpha \widetilde{F}_{B \mu \nu} \right) f^{\mu \nu}
- \left( \partial^{\mu} H_{B \mu \nu} \right) f^{\nu\alpha} \, .
\end{eqnarray}
Now, we multiply the axion field equation (\ref{eqescalar}) by $\partial^{\alpha} \widetilde{\phi}$ and, after some algebraic manipulations,
we end up with
\begin{eqnarray} \label{eq_T2}
\partial^\mu \left\{  \partial_\mu \widetilde{\phi} \, \partial^\alpha \widetilde{\phi}
- \delta_{\mu}^{\;\;\,\alpha} \left[ \,  \frac{1}{2} \, ( \partial_\nu \widetilde{\phi} )^2 - \frac{1}{2} \, m^2 \, \widetilde{\phi}^2
\, \right] \right\}
=- \frac{1}{2} \, g \, \widetilde{F}_{B\mu\nu} \, f^{\mu \nu} \,  \partial^\alpha \widetilde{\phi} \; .
\end{eqnarray}
The sum of the equations \eqref{eq_T1} with \eqref{eq_T2} yields the result
\begin{eqnarray}\label{EqThetatotal}
\partial^{\mu} \, \Theta_{\mu}^{\; \; \, \alpha} = \Omega^{\alpha}  \; ,
\end{eqnarray}
where the energy-momentum tensor of the system is given by
\begin{eqnarray} \label{Tmunu}
 \Theta_{\mu}^{\; \; \, \alpha} &=&  ( \partial_{\mu} \widetilde{\phi} ) \, ( \partial^{\alpha}\widetilde{\phi} )
+c_1 \, f_{\mu\nu} \, f^{\nu\alpha} - \frac{1}{2} \, Q_{B\mu\nu\kappa\lambda} \, f^{\kappa\lambda} \,  f^{\nu\alpha}
+g \, \widetilde{\phi} \, \widetilde{F}_{B \mu\nu} \, f^{\nu\alpha}
\nonumber \\
&&
\hspace{-0.6cm}
-\,\delta_{ \mu}^{\;\;\, \alpha} \left[ \, -\frac{1}{4} \, c_{1} \, f_{\rho\sigma}^{\, 2}  \right.
+\left. \frac{1}{8} \, Q_{B\rho\sigma\kappa\lambda}f^{\rho\sigma}f^{\kappa\lambda} + \frac{1}{2} \, ( \partial_{\rho}\widetilde{\phi} )^{2}
-\frac{1}{2} \, m^2 \, \widetilde{\phi}^2
-\frac{1}{2} \, g \, \widetilde{\phi} \, \widetilde{F}_{B\rho\sigma} \, f^{\rho\sigma} \, \right] \, , \;
\end{eqnarray}
and $\Omega_{\alpha}$ corresponds to the dissipative terms, namely,
\begin{eqnarray}
 \Omega^{\alpha} &=&  J_{\nu} \, f^{\nu\alpha}
+ \frac{1}{2} \, g \, \widetilde{\phi} \, \left( \partial^{\alpha}  \widetilde{F}_{B\mu\nu} \right) \, f^{\mu\nu}
-\left( \partial^{\mu} H_{B \mu \nu} \right) f^{\nu\alpha}
\nonumber \\
&+& \! \frac{1}{4} \left(\partial^{\alpha} c_{1} \right) f_{\mu\nu}^{\, 2}
+\frac{1}{4} \left(\partial^{\alpha}c_{2} \right) \widetilde{f}_{\mu\nu}f^{\mu\nu}
-\frac{1}{8} \, \left(\partial^{\alpha} Q_{B\mu\nu\kappa\lambda}\right)f^{\mu\nu}f^{\kappa\lambda}
\; .
\end{eqnarray}
The energy-momentum tensor \eqref{Tmunu} exhibits the well-known contributions of non-linear electrodynamics and axion field. Furthermore, due the presence of an electromagnetic background, we also obtain axion-photon mixing terms related to $\widetilde{\phi} \, \widetilde{F}_{B \mu\nu} \, f^{\nu\alpha}$ and
$\widetilde{\phi} \, \delta_{ \mu}^{\;\; \alpha} \, \widetilde{F}_{B\rho\sigma} \, f^{\rho\sigma}$. We highlight that our result does not include the influence of magnetic monopoles because we used the Bianchi identity. Notice that $\Omega^{\alpha}$ contains the usual contribution $J_{\nu} \, f^{\nu\alpha}$ and other terms involving derivatives of the background fields. For the particular case of vanishing axion field, we recover the results discussed in ref. \cite{MJNevesEDN}. We point out that the energy-momentum tensor is not symmetric, $\Theta^{\mu\alpha} \neq \Theta^{\alpha\mu}$, due to the background terms $Q_{B\mu\nu\kappa\lambda} \, f^{\kappa\lambda} \, f^{\nu\alpha}$
and $g \, \widetilde{\phi} \, \widetilde{F}_{B \mu\nu} \, f^{\nu\alpha}$. This fact also happens in scenarios with Lorentz symmetry violation  (see, for instance, refs. \cite{Bonetti,Spallicci}). If we consider a uniform and constant electromagnetic background with $J^{\mu}=0$, the eq. (\ref{EqThetatotal}) leads to the conservation law $\partial^{\mu} \Theta_{\mu}^{\; \; \alpha}=0$. When $\alpha=0$, we obtain
\begin{equation} \label{law_u_S}
\partial_{t}u+\nabla\cdot{\bf S}=0 \, ,
\end{equation}
where $u:=\Theta^{00}$ denotes the energy density,
\begin{eqnarray}\label{theta00}
u &=& \frac{1}{2} \, (\partial_{t}\widetilde{\phi})^{2}
+ \frac{1}{2} \, (\nabla\widetilde{\phi})^{2}+ \frac{1}{2} \, m^2 \, \widetilde{\phi}^2
+ \frac{1}{2} \, c_{1} \left( {\bf e}^2+{\bf b}^2 \right)
\nonumber \\
&&
\hspace{-0.5cm}
+ \frac{1}{2} \, d_1 \, \left( {\bf e}\cdot{\bf E} \right)^2
+ \frac{1}{2} \, d_2 \, \left( {\bf e}\cdot{\bf B} \right)^2
-\frac{1}{2} \, d_{1} \left( {\bf b}\cdot{\bf B} \right)^2
\nonumber \\
&&
\hspace{-0.5cm}
-\frac{1}{2} \, d_2 \left( {\bf b}\cdot{\bf E} \right)^2
+d_3 \left( {\bf e}\cdot{\bf E} \right) \left({\bf e}\cdot{\bf B}\right)
+d_{3} \left( {\bf b}\cdot{\bf E} \right)\left( {\bf b}\cdot{\bf B} \right)
\nonumber \\
&&
\hspace{-0.5cm}
+g \, \widetilde{\phi} \, ({\bf b}\cdot{\bf E})+g \, \widetilde{\phi} \, ({\bf e}\cdot{\bf B}) \, ,
\label{theta00}
\end{eqnarray}
and ${\bf S}$ corresponds to the Poynting vector, whose components are defined by $S^{i}:=\Theta^{i0}$, such that
\begin{eqnarray}\label{Si}
S^{i} &=& (\partial^i \widetilde{\phi}) \; (\partial_t \widetilde{\phi})
+ c_1 ({\bf e} \times {\bf b} )^i + d_1 \, ({\bf e} \cdot {\bf E})\;({\bf e} \times {\bf B} )^i
\nonumber \\
&&
\hspace{-0.5cm}
-d_1 \, ({\bf b} \cdot {\bf B})\;({\bf e} \times {\bf B} )^i
-d_2 \, ({\bf e} \cdot {\bf B})\;({\bf e} \times {\bf E} )^i
-d_2 \, ({\bf b} \cdot {\bf E})\;({\bf e} \times {\bf E} )^i
\nonumber \\
&&
\hspace{-0.5cm}
+d_3 \, ({\bf e} \cdot {\bf B})\;({\bf e} \times {\bf B} )^i
-d_3 \, ({\bf e} \cdot {\bf E})\;({\bf e} \times {\bf E} )^i
+d_3 \, ({\bf b} \cdot {\bf E})\;({\bf e} \times {\bf B} )^i
\nonumber \\
&&
\hspace{-0.5cm}
-d_3 \, ({\bf b} \cdot {\bf B})\;({\bf e} \times {\bf B} )^i
-g \, \widetilde{\phi} \left({\bf e} \times {\bf E} \right)^i \; .
\label{theta0i}
\end{eqnarray}
For $\alpha=j$, the conservation law  is written as
\begin{equation} \label{law_P_T}
\partial_{t}{\bf P}+\nabla\cdot \overleftrightarrow{{\bf T}}={\bf 0} \, ,
\end{equation}
where ${\bf P}$ stands for the momentum, with the components $P^{i}:=\Theta^{0i}$ given by
\begin{eqnarray}\label{Pi}
P^{i} &=& (\partial_{t}\widetilde{\phi}) \, (\partial^{\,i}\widetilde{\phi}) + c_{1} \left( {\bf e}\times{\bf b} \right)^{i}
-d_1 \, \left( {\bf e} \cdot {\bf E} \right) \left( {\bf E}\times{\bf b} \right)^{i}
\nonumber \\
&&
\hspace{-0.5cm}
-d_1 \, \left( {\bf b} \cdot {\bf B} \right) \left( {\bf E}\times{\bf b} \right)^{i}
-d_2 \, \left( {\bf e} \cdot {\bf B} \right) \left( {\bf B}\times{\bf b} \right)^{i}
\nonumber \\
&&
\hspace{-0.5cm}
+d_2 \, \left( {\bf b} \cdot {\bf E} \right) \left( {\bf B}\times{\bf b} \right)^{i}
-d_3 \, \left( {\bf e} \cdot {\bf B} \right) \left( {\bf E}\times{\bf b} \right)^{i}
\nonumber \\
&&
\hspace{-0.5cm}
-d_3 \, \left( {\bf e} \cdot {\bf E} \right) \left( {\bf B}\times{\bf b} \right)^{i}
-d_3 \, \left( {\bf b} \cdot {\bf B} \right) \left( {\bf B}\times{\bf b} \right)^{i}
\nonumber \\
&&
\hspace{-0.5cm}
+d_3 \, \left( {\bf b} \cdot {\bf E} \right) \left( {\bf E}\times{\bf b} \right)^{i}
+g \, \widetilde{\phi} \, \left( {\bf B} \times {\bf b} \right)^{i} \; ,
\end{eqnarray}
and $\overleftrightarrow{{\bf T}}$ denotes the stress tensor, whose components
$(\overleftrightarrow{{\bf T}})^{ij}:=\Theta^{ij}$ yield the expression
\begin{eqnarray}
(\overleftrightarrow{{\bf T}})^{ij} &=& (\partial^i \widetilde{\phi}) \; (\partial^j \widetilde{\phi}) - c_1 (e^i \, e^j+b^i \, b^j) -d_1({\bf e} \cdot {\bf E})\;E^i \,e^j - d_2({\bf e} \cdot {\bf B})\;B^i \, e^j
\nonumber \\
&&
\hspace{-0.5cm}
+d_1({\bf b} \cdot {\bf B})\;E^i \, e^j - d_2({\bf b} \cdot {\bf E})\;B^i \, e^j
-d_1({\bf e} \cdot {\bf E})\;b^i \, B^j + d_2({\bf e} \cdot {\bf B})\;b^i \, E^j
\nonumber \\
&&
\hspace{-0.5cm}
+ d_1({\bf b} \cdot {\bf B})\;b^i \, B^j + d_2({\bf b} \cdot {\bf E})\;b^i \, E^j
-d_3({\bf e} \cdot {\bf B})\;E^i \, e^j - d_3({\bf e} \cdot {\bf E})\;B^i \, e^j
\nonumber \\
&&
\hspace{-0.5cm}
-d_3({\bf b} \cdot {\bf E})\;E^i \, e^j + d_3({\bf b} \cdot {\bf B})\;B^i \, e^j
- d_3({\bf e} \cdot {\bf B})\;b^i \, B^j + d_3({\bf e} \cdot {\bf E})\;b^i \, E^j
\nonumber \\
&&
\hspace{-0.5cm}
-d_3({\bf b} \cdot {\bf E})\;b^i \, B^j - d_3({\bf b} \cdot {\bf B})\;b^i \, E^j
+g \, \widetilde{\phi} \, (b^i \, E^j - B^i \, e^j) \; .
\end{eqnarray}
The presence of an electromagnetic background introduces some interesting features. Firstly,
the energy density of the model acquires new contributions involving the axion field, given by $g \, \widetilde{\phi} \, ({\bf b}\cdot{\bf E})$ and $g \, \widetilde{\phi} \, ({\bf e}\cdot{\bf B})$. In addition, from the eqs. (\ref{Si}) and (\ref{Pi}), we conclude that the Poynting vector does not coincide with the linear momentum. However, as already expected, both equations lead to the same expression if the electromagnetic background is switched of.
Finally, it should be mentioned that using the above definitions for $u, {\bf S}, {\bf P}$ and $\overleftrightarrow{{\bf T}}$, based on the conservation laws (\ref{law_u_S}) and (\ref{law_P_T}), the results will be consistent.


\section*{Acknowledgements}
The work of M. J. Neves has been supported by the Conselho Nacional de Desenvolvimento Cient\'ifico e Tecnol\'ogico (CNPq) under grant 313467/2018-8 (GM).  L. P. R. Ospedal is grateful to CNPq for his Post-Doctoral Fellowship. He thanks the COSMO-CBPF group for their friendly hospitality when part of this work was developed. The work of J. M. A. Paix\~ao has been supported by CNPq.
The authors also express their gratitude to P. Gaete and A. Spallicci
for long and stimulating discussions.
%

%


%

\begin{thebibliography} {99}
%


\bibitem{Peccei}  R. D. Peccei and Helen R. Quinn, {\it $\mathrm{CP}$ Conservation in the Presence of Pseudoparticles}, Phys. Rev. Lett. {\bf 38} (1977) 1440.


\bibitem{Peccei2}  R. D. Peccei and Helen R. Quinn, {\it Constraints imposed by $\mathrm{CP}$ conservation in the presence of pseudoparticles}, Phys. Rev. D {\bf 16} (1977) 1791.


\bibitem{Wilczekaxion}  J. Preskill, M. B. Wise and F. Wilczek, {\it Cosmology of the invisible axion}, Phys. Lett. B {\bf 120} (1983) 127.


\bibitem{Sikivie}  L. F. Abbott and P. Sikivie, {\it A cosmological bound on the invisible axion}, Phys. Lett. B {\bf 120} (1983) 133.


\bibitem{Dine}  M. Dine and W. Fischler, {\it The not-so-harmless axion}, Phys. Lett. B {\bf 120} (1983) 137.


\bibitem{Ayala}  Adrian Ayala,  Inma Dom\'{\i}nguez,  Maurizio Giannotti,  Alessandro Mirizzi and Oscar Straniero, {\it Revisiting the Bound on Axion-Photon Coupling from Globular Clusters}, Phys. Rev. Lett. {\bf 113} (2014) 191302 [arXiv:hep-ph/1406.6053].


\bibitem{Reynolds} Christopher S. Reynolds, M. C. David Marsh, Helen R. Russell, Andrew C. Fabian, Robyn Smith, Francesco Tombesi and Sylvain Veilleux, {\it Astrophysical Limits on Very Light Axion-like Particles from Chandra Grating Spectroscopy of {NGC} 1275}, The Astrophysical Journal {\bf 890} (2020) 59 [arXiv:hep-ph/1907.05475v4].

\bibitem{Bondarenko} Kyrylo Bondarenko, Alexey Boyarsky, Josef Pradler and  Anastasia Sokolenko, {\it Neutron stars as photon double-lenses: constraining resonant conversion into ALPs}, [arXiv:hep-ph/2203.08663].

\bibitem{CAST} The CAST Collaboration, {\it New CAST limit on the axion-photon interaction}, Nature Physics {\bf 12} (2017) 584 [arXiv:hep-ex/1705.02290v2].


\bibitem{SCHOEFFEL} L. Schoeffel, C. Baldenegro, H. Hamdaoui, S. Hassani, C. Royon and M. Saimpert, {\it Photon photon physics at the LHC and laser beam experiments, present and future}, Progress in Particle and Nuclear Physics {\bf 120} (2021) 103889 [arXiv:hep-ph/2010.07855v3].


\bibitem{Baldenegro1}  Cristian Baldenegro, Sylvain Fichet, Gero von Gersdorff and Christophe Royon, {\it Searching for axion-like particles with proton tagging at the LHC}, JHEP {\bf 2018} (2018) 1  [arXiv:hep-ph/1803.10835v1].


\bibitem{Baldenegro2}  C. Baldenegro,  S. Hassani,  C. Royon, L. Schoeffel, {\it Extending the constraint for axion-like particles as resonances at the LHC and laser beam experiments}, Phys. Lett. B {\bf 795} (2019) 339 [arXiv:hep-th/1903.04151v4].

\bibitem{d'Enterria} David d'Enterria, {\it Collider constraints on axion-like particles}, Particles (FIPs) 2020 Workshop Report, (2021) [arXiv:hep-ex/2102.08971v2].




\bibitem{Schwinger} Julian Schwinger, {\it On Gauge Invariance and Vacuum Polarization}, Phys. Rev. {\bf 82} (1951) 664.


\bibitem{Duncan}  Robert C. Duncan and Christopher Thompson, {\it Formation of very strongly magnetized neutron stars-Implications for gamma-ray bursts}, The Astrophysical Journal {\bf 392} (1992) L9.



\bibitem{jacobsen2022constraining} Sunniva Jacobsen, Tim Linden and Katherine Freese, {\it Constraining Axion-Like Particles with HAWC Observations of TeV Blazars}, [arXiv:hep-ph/2203.04332v1].


\bibitem{2022nonlinear} Ayd{\i}n Cem Keser, Yuli Lyanda-Geller and Oleg P. Sushkov, {\it Nonlinear Quantum Electrodynamics in Dirac Materials}, Phys. Rev. Lett. {\bf 128} (2022) 66402 [arXiv:cond-mat.other/2101.09714v2].


\bibitem{Plebanski} J. Plebanski, {\it Lectures on nonlinear electrodynamics},
Cycle of lectures delivered at The Niels Bohr Institute and NORDITA, Copenhagen, October 1968.

\bibitem{Boillat} G. Boillat, {\it Nonlinear electrodynamics: Lagrangian and equations of motion},
J. Math. Phys. {\bf 11} (1970) 941.

\bibitem{Birula1} Z. Bialynicka-Birula and I. Bialynicki-Birula, {\it Nonlinear effects in Quantum Electrodynamics. Photon propagation and photon splitting in an external field}, Phys. Rev. D {\bf 2}  (1970) 2341.

\bibitem{Birula2} I. Bialynicki-Birula, {\it Nonlinear Electrodynamics: Variations on a theme by Born and Infeld}, in Quantum Theory of Particles and Fields: birthday volume dedicated to Jan Lopusza\'nski, eds B. Jancewicz and J. Lukierski, World Scientific, 1983, pp 31.

\bibitem{Sorokin} Dmitri P. Sorokin, {\it Introductory Notes on Non-linear Electrodynamics and its Applications}, [arXiv:hep-th/2112.12118v2].


\bibitem{TTdeformations1} H. Babaei-Aghbolagh, K.B Velni, D.M. Yekta and H. Mohammadzadeh, {\it Emergence of non-linear electrodynamic theories from $ T\overline{T} $-like deformations}, Phys. Lett. B {\bf 829} (2022) 137079 [arXiv:hep-th/2202.11156].


\bibitem{TTdeformations2} H. Babaei-Aghbolagh, K.B Velni, D.M. Yekta and H. Mohammadzadeh, {\it $ T\overline{T} $-like flows in non-linear electrodynamic theories and S-duality}, JHEP {\bf 2021} (2021) 1 [arXiv:hep-th/2012.13636v3].



\bibitem{Born} M. Born and L. Infeld, {\it Foundations of the new field theory}, Proc. R. Soc. Lond. Ser. A {\bf 144} (1934) 425.


\bibitem{colegarussa} E. M. Murchikova, {\it On nonlinear classical electrodynamics with an axionic term},  J. Phys. A: Math. Theor. {\bf 44} (2011) 045401 [arXiv:cond-mat.mes-hall/1005.5027v3].



\bibitem{Fradkin} E. S. Fradkin and A. A. Tseytlin, {\it Non-linear electrodynamics from quantized strings},
Phys. Lett. B {\bf 163} (1985) 123.


\bibitem{Pope} E. Bergshoeff, E. Sezgin, C. N. Pope and P. K. Townsend, {\it The Born-Infeld action from conformal invariance of the open superstring},
Phys. Lett. B {\bf 188} (1987) 70 .


\bibitem{Banerjee} R. Banerjee and D. Roychowdhury, {\it Critical behavior of Born-Infeld AdS black holes in higher dimensions},
Phys. Rev. D {\bf 85} (2012) 104043 [arXiv:gr-qc/1203.0118v2].


\bibitem{Mann} S. Gunasekaran, D. Kubiznak and R. B. Mann, {\it Extended phase space thermodynamics for charged and rotating black holes and Born-Infeld vacuum polarization}, JHEP {\bf 1211} (2012) 110 [arXiv:hep-th/1208.6251v2].


\bibitem{Garcia} Eloy Ay\'on-Beato and Alberto Garc\'ia {\it The Bardeen Model as a Nonlinear Magnetic Monopole}, Phys. Lett. B {\bf 493} (2000) 149-152 [arXiv:gr-qc/0009077v1].


\bibitem{NiauEPJC} P. Niau Akmansoy and L. G. Medeiros, {\it Constraining Born-Infeld-like nonlinear
electrodynamics using hydrogen's ionization energy}, Eur. Phys. J. C {\bf 78}
(2018) 143 [arXiv:hep-ph/1712.05486].


\bibitem{NiauPRD} P. Niau Akmansoy and L. G. Medeiros, {\it Constraining nonlinear corrections to
Maxwell electrodynamics using $\gamma\gamma$ scattering}, Phys. Rev. D {\bf 99}
(2019) 115005 [arXiv:hep-ph/1809.01296].


\bibitem{Ellis} John Ellis, Nick E. Mavromatos and Tevong You, {\it Light-by-Light Scattering Constraint on Born-Infeld Theory}, Phys. Rev. Lett. {\bf 118} (2017) 261802 [arXiv:hep-ph/1703.08450v2].


\bibitem{MJNevesPat} M. J. Neves,  L. P. R. Ospedal,  J. A. Helay\"el-Neto and Patricio Gaete, {\it Considerations on anomalous photon and $Z$-boson self-couplings from the Born-Infeld weak hypercharge action}, Eur. Phys. J. C {\bf 82} (2022) 327 [arXiv:hep-th/2109.11004v2].


\bibitem{Battesti} R. Battesti and C. Rizzo, {\it Magnetic and electric properties of a quantum vacuum}, Rep. Prog. Phys. {\bf 76} (2013) 016401 [arXiv:physics.optics/1211.1933v1].


\bibitem{EulerHeisen} W. Heisenberg and H. Euler, {\it Consequences of dirac theory of the positron}, Z. Phys. {\bf 98} (1936) 714.


\bibitem{Weisskopf} V. S. Weisskopf, {\it {\"U}ber die elektrodynamik des vakuums auf grund des quanten-theorie des elektrons},
Dan. Mat. Fys. Medd. {\bf 14} (1936) 1.


\bibitem{25years} A. Ejlli, F. Della Valle, U. Gastaldi, G. Messineo, R. Pengo, G. Ruoso and G. Zavattini, {\it The PVLAS experiment: A 25 year effort to measure vacuum magnetic birefringence},
Phys. Rep. {\bf 871} (2020) 1 [arXiv:hep-ex/2005.12913v1].


\bibitem{Zavattini} G. Zavattini, U. Gastaldi, R. Pengo, G, Ruoso, F Della Valle and E. Milotti, {\it Measuring the magnetic birefringence of vacuum: the PVLAS experiment}, Int. J. of Mod. Phys. A {\bf 27} (2012) 1260017 [arXiv:hep-ex/arXiv:1201.2309v1].


\bibitem{DellaValle} F. Della Valle, A. Ejlli, U. Gastaldi, G. Messineo, E. Milotti, R. Pengo, G. Ruoso and G. Zavattini, {\it The PVLAS experiment: measuring vacuum magnetic birefringence and dichroism with a birefringent Fabry--Perot cavity}, Eur. Phys. J. C {\bf 76} (2016) 1 [arXiv:physics.optics/1510.08052].


\bibitem{Zavattini_Universe} G. Zavattini and F. Della Valle, {\it Optical polarimetry for fundamental physics}, Universe {\bf 7} (2021) 252 


\bibitem{Mignani} R. P. Mignani, V. Testa, D. Gonzalez Caniulef, R. Taverna, R. Turolla, S. Zane and K. Wu, {\it Evidence for vacuum birefringence from the first optical polarimetry measurement of the isolated neutron star RX J1856. 5- 3754}, Monthly Notices of the Royal Astronomical Society (2016) stw2798.


\bibitem{Yamazaki} T. Yamazaki, T. Inada, T. Namba, S. Asai, T. Kobayashi, A. Matsuo, K. Kindo and H. Nojiri, {\it Repeating pulsed magnet system for axion-like particle searches and vacuum birefringence experiments}, Nuclear Instruments and Methods in Physics Research Section A: Accelerators, Spectrometers, Detectors and Associated Equipment {\bf 833} (2016) 122 [arXiv:physics.ins-det/1604.06556v3].


\bibitem{Maiani}  L. Maiani, R. Petronzio and E. Zavattini, {\it Effects of nearly massless, spin-zero particles on light propagation in a magnetic field}, Phys. Lett. B {\bf 175} (1986) 359.


\bibitem{Villalba} S. Villalba-Ch{\'a}vez and Antonino Di Piazza, {\it Axion-induced birefringence effects in laser driven nonlinear vacuum interaction},  JHEP {\bf 11} (2013) 136 [arXiv:hep-ph/1307.7935v2].


\bibitem{Scott} Scott Robertson, {\it Optical Kerr effect in vacuum}, Phys. Rev. A {\bf 100}, 063831 (2019) [arXiv:physics-optics/1908.00896v3].


\bibitem{Rizzo} G. L. J. A. Rikken and C. Rizzo, {\it Magnetoelectric birefringences of the quantum vacuum}, Phys. Rev. A {\bf 63}, 012107 (2000).





\bibitem{Raffelt_1988}  Georg Raffelt and Leo Stodolsky, {\it Mixing of the photon with low-mass particles}, Phys. Rev. D {\bf 37} (1988).  


\bibitem{Gaete_1} Patricio Gaete, {\it  Some Considerations About Podolsky-Axionic Electrodynamics}, Int. J. Mod. Phys. A {\bf 27} (2012) 1250061 [arXiv:hep-th/1110.4816].


\bibitem{Gaete_2} Patricio Gaete and Iv\'an Schmidt, {\it Properties of noncommutative axionic electrodynamics}, Phys. Rev. D {\bf 76}  (2007) 027702 [arXiv:hep-th/0706.3176].

\bibitem{Gaete_3} Patricio Gaete and Euro Spallucci, {\it  Finite axionic electrodynamics from a new noncommutative approach}, J. Phys. A {\bf 45}  (2012) 065401 [arXiv:hep-th/1102.3777v3].

\bibitem{Li_Cheng_2008} Cheng-De Li and Yan-Fu Cheng, {\it Proca Effects of Axion-Like Particle-Photon Interactions
on Light Polarization in External Magnetic Fields}, Int. J. Theor. Phys. {\bf 47},  (2008) 1911.


\bibitem{Chrispim} B. A. S. D. Chrispim, R. C. L. Bruni and M. S. Guimaraes, {\it Massive photon propagator in the presence of axionic fluctuations}, Phys. Rev. B {\bf 103} (2021) 165120 [arXiv:hep-ph/2012.00184v3].



\bibitem{Huang_Lee} Fa Peng Huang and Hye-Sung Lee, {\it Extension of the electrodynamics in the presence of the axion and dark photon}, Int. J. Mod. Phys. A {\bf 34} (2019) 1950012 [arXiv:hep-ph/1806.09972v3].
%

\bibitem{Arias} Paola Arias, Ariel Arza, Joerg Jaeckel and Diego Vargas-Arancibia, {\it Hidden Photon Dark Matter Interacting via Axion-like Particles},  JCAP {\bf 05} (2021) 070 [arXiv:hep-ph/2007.12585v2].
%

\bibitem{PedroSilvaPRD2021} Pedro D. S. Silva, Let{\'i}cia Lisboa-Santos, Manoel M. Ferreira Jr. and Marco Schreck, {\it Effects of CPT-odd terms of dimensions three and five on electromagnetic propagation in continuous matter}, Phys. Rev. D {\bf 104}, 116023 (2021) [arXiv:hep-th/2109.04659v2].


\bibitem{PedroSilvaArxiv2022} Pedro D. S. Silva, Rodolfo Casana, Manoel M. Ferreira Jr, {\it Symmetric and antisymmetric constitutive tensors for bi-isotropic and bi-anisotropic media}, arXiv:physics.class-ph/2204.10460v1.


\bibitem{BorgesPRD2014} L. H. C. Borges, A. G. Dias, A. F. Ferrari, J. R. Nascimento, and A. Yu. Petrov, {\it Generation of axionlike couplings via quantum corrections in a Lorentz-violating background}, Phys. Rev. D {\bf 89} 045005 [arXiv:hep-th/1304.5484v2].


\bibitem{Pierre} Pierre Sikivie, {\it Invisible Axion Search Methods}, Rev. Mod. Phys. {\bf 93} (2021) 15004  [arXiv:hep-ph/2003.02206v2].
%

\bibitem{PDG} P. A. Zyla et al. (Particle Data Group), Prog. Theor. Exp. Phys. 2020, 083C01 (2020) and 2021 update. See section $90$: {\it  Axions and Other Similar Particles}.


\bibitem{livroCaloz} Christophe Caloz and Tatsuo Itoh, {\it Electromagnetic metamaterials: Transmission line theory and microwave applications}, A John Wiley $\&$ sons, INC., publication (2006), pg. 03.

\bibitem{Veselago} V. Veselago, {\it The electrodynamics of substances with simultaneously negative  permissibility and permittivity}, Sov. Phys. Uspekhi {\bf 10} (1968) 50.

\bibitem{Padilla} W. J. Padilla, D. N. Basov, D. R. Smith, {\it Negative refractive index metamaterials}, Materials today {\bf 9} (2006) 28.


\bibitem{MJNevesEDN} M. J. Neves, Jorge B. de Oliveira, L. P. R. Ospedal and J. A. Helay\"el-Neto, {\it Dispersion Relations in Non-Linear Electrodynamics and the Kinematics of the Compton Effect in a Magnetic Background},  Phys. Rev. D {\bf 104} (2021) 015006 [arXiv:hep-th/2101.03642v3].



\bibitem{Tobar_PRD} Michael E. Tobar, Ben T. McAllister and Maxim Goryachev, {\it Poynting vector controversy in axion modified electrodynamics}, Phys. Rev. D {\bf 105} (2022) 045009 [arXiv:hep-th/2109.04056v4].

\bibitem{Zhou} Kevin Zhou, {\it Comment on "Poynting vector controversy in axion modified electrodynamics"}, [arXiv:hep-th/2203.15821v1].

\bibitem{Patkos} Andr{\'a}s Patk{\'o}s, {\it Radiation backreaction in axion electrodynamics}, Symmetry {\bf 14} (2022) 1113  [arXiv:hep-ph/2206.02052v1].


\bibitem{Bonetti} Luca Bonetti, Lu\'is R. dos Santos Filho, Jos\'e A. Helay\"el-Neto and Alessandro D. A. M. Spallicci, {\it Photon sector analysis of Super and Lorentz symmetry breaking: effective photon mass, bi-refringence and dissipation}, Eur. Phys. J. C {\bf 78} (2018) 811 [arXiv:hep-th/1709.04995v2].

\bibitem{Spallicci} Jos\'e A. Helay\"el-Neto and Alessandro D. A. M. Spallicci, {\it Frequency variation for in vacuo photon propagation in the Standard-Model Extension}, Eur. Phys. J. C {\bf 79} (2019) 590 [arXiv:hep-th/1904.11035v2].

%
%
\end{thebibliography}
\end{document}